\documentclass[5p,preprint]{elsarticle}
\biboptions{sort&compress}
\usepackage{lineno,isotope,mathrsfs}
\modulolinenumbers[5]
\usepackage{amsmath,amssymb,mathptmx,epsfig,bm,mathrsfs}
\usepackage{color,slashed,nicefrac,exscale,multirow,soul}
\usepackage{times,txfonts,xcolor,graphicx,gensymb,tikz,booktabs}
\usepackage[normalem]{ulem}
\usepackage[breaklinks,colorlinks,citecolor=blue]{hyperref}
\newcommand{\sat}[1]{{#1}_{\mathrm{sat}}}
\newcommand{\sym}[1]{{#1}_{\mathrm{sym}}}
\newcommand{\tth}[1]{{#1}_{\mathrm{th}}}
\newcommand{\be}{\begin{equation}}
\newcommand{\ee}{\end{equation}}
\newcommand{\bea}{\begin{eqnarray}}
\newcommand{\eea}{\end{eqnarray}}
\begin{document}
\begin{frontmatter}
  \title{Bayesian inference of thermal effects in dense matter within
  the covariant density functional theory}
  
\author[a]{Adriana R. Raduta}
\ead{araduta@nipne.ro}

\author[a]{Mikhail V. Beznogov}
\ead{mikhail.beznogov@nipne.ro}

\author[b]{Micaela Oertel}
\ead{micaela.oertel@obspm.fr}

\address[a]{National Institute for Physics and Nuclear Engineering (IFIN-HH), RO-077125 Bucharest, Romania}

\address[b]{LUTH, Observatoire de Paris, Universit\'e PSL, CNRS,
  Universit\'e de Paris Cité, 92190 Meudon, France}

\date{\today}
\begin{abstract}
 The high temperatures reached in a proto-neutron star or during the post-merger phase of a binary neutron star coalescence lead to non-negligible thermal effects on the equation of state (EOS) of dense nuclear matter (NM). Here we study these effects within the covariant density functional theory employing the posteriors of a Bayesian inference, which encompasses a large sample of EOS models. Different densities and  temperatures are considered. We find that for a number of quantities thermal effects are strongly correlated with the Dirac effective mass ($m^*$) of the nucleons and/or its logarithmic derivative as a function of density. These results can be explained within the low temperature approximation though they survive beyond this limit.
\end{abstract}
\begin{keyword}
Equation of state  \sep Hot and dense matter \sep Thermal effects \sep Covariant density functional
\end{keyword}
\end{frontmatter}

\section{Introduction}
\label{sec:Intro}

Numerical simulations of core-collapse supernovae and the subsequent evolution of the proto-neutron star into a mature cold neutron star (NS), simulations of the formation of stellar black holes and of binary neutron star (BNS) mergers require detailed knowledge of the equation of state (EOS) over wide domains of baryon number density $10^{-14}~\mathrm{fm}^{-3} \lesssim n_\mathrm{B} \lesssim 2 ~\mathrm{fm}^{-3}$, temperature $0 \leq T \lesssim 100~\mathrm{MeV}$ and electron fraction $0 \leq Y_{\mathrm e}\lesssim0.6$~\cite{Oertel_RMP_2017,Burgio_Fantina2018}.

More than a hundred different EOS models considering non-zero temperatures have been developed so far, see, e.g., those available in tabulated form on \textsc{CompOSE}~\cite{Typel_2013,Typel_EPJA_2022}\footnote{\url{https://compose.obspm.fr/}}. The huge majority of them rely either on a non-relativistic potential model or on the covariant density functional (CDF) theory to describe dense nuclear matter (NM). They assume that the effective interactions have the same functional form at zero and finite temperatures.

In default of constraints from nuclear experiments, and awaiting third generation gravitational wave detectors which might be able to constrain thermal effects in the post-merger phase of a BNS, see, e.g., Ref.~\cite{Raithel:2023gct}, insights into the thermal behavior of NM can be gained by confronting the predictions of available models against each other and by comparison with ab initio calculations. Pioneering works of Constantinou et al.~\cite{Constantinou_PRC_2014,Constantinou_PRC_2015} have proved that, in non-relativistic potential models, the thermal response functions are controlled by the Landau effective masses of the nucleons and their derivatives with respect to density. This feature has, for instance, been used to assess the importance of thermal effects in the post-merger remnant in Ref.~\cite{Fields_ApJL_2023}.

Models with finite range interactions have been further shown to present traits distinct from those obtained by models with zero-range interactions~\cite{Constantinou_PRC_2015,Constantinou_Annals_2015}. Non-relativistic models with finite- and zero-range interactions differ in the predictions for the specific heat at constant volume ($C_\mathrm{V}$) and the thermal index ($\tth{\Gamma}$) as functions of density. While Skyrme models predict for $C_\mathrm{V}$ a behavior similar to the one of free Fermi gases, where the maximum value of 1.5 is obtained for vanishing densities, non-relativistic models with more involved momentum dependencies lead to a non-monotonic behavior of $C_\mathrm{V}$ as a function of density, with a maximum, in excess of 1.5, occurring at finite density. Besides, in standard Skyrme models $\tth{\Gamma}$ has a strong and monotonic increase with the density, while in models with finite range interactions it is almost flat. All these differences originate from the differences in the density dependence of effective masses which, for finite-range interactions, also depend on temperature~\cite{Constantinou_Annals_2015}.
Qualitative differences among non-relativistic and CDF-based models are obvious in the density dependence of thermal contributions to pressure ($p$) and chemical potential ($\mu$) as well as in the density dependence of the thermal index. The thermal contribution to a state variable is defined as the difference between the values that the quantity takes at finite and zero temperatures, i.e. $\tth{X}=X(T,n,Y_{\mathrm p})-X(0,n,Y_{\mathrm p})$, where $n$ is the density and $Y_{\mathrm p}=Y_{\mathrm e}$ is the proton fraction. The thermal index, $\tth{\Gamma}=1+\tth{p}/\tth{e}$, quantifies the departure from the ideal gas behavior ($\tth{e}$ stands for the thermal energy density).
Non-relativistic models predict for $\tth{p}$ and $\tth{\mu}$ a steep (zero-range interactions) or gentle (finite-range interactions) increase with the density, while in CDF models these two quantities reach their maxima at finite densities and then saturate. The latter feature entails a strong decrease of $\tth{\Gamma}$ at high densities. Analytic formulas derived at next-to-leading order in an expansion in temperature have further shown that, for all these types of models, temperature corrections to thermal state variables can be expressed in terms of the level density parameter, Fermi momentum, effective mass and its derivative with respect to density~\cite{Constantinou_Annals_2015}. The sophistication degree of the model gets reflected into the complexity of these formulas. The simplest expressions correspond to Skyrme interactions.

Ref.~\cite{Raduta_EPJA_2021} systematically confronted the thermal behavior of a selection of mean field models frequently used in astrophysical numerical simulations. It came out that constraining a Skyrme potential to reproduce the density dependence of the effective mass in ab initio models results in a thermal behavior deviating from the one of standard models, see, e.g., the high density increase of $\tth{e}$ and related decrease of $\tth{\Gamma}$ of NRAPR~\cite{Steiner_PhysRep_2005} in Ref.~\cite{Raduta_EPJA_2021}. On the other hand, the thermal response of CDF models has been shown to strongly depend on the mesonic couplings, with $\tth{p}$ being the most affected quantity.
The role of non-nucleonic degrees of freedom has been further investigated in Ref.~\cite{Raduta_EPJA_2022},
whose most preeminent results are the increase (decrease) of thermal energy density (pressure) with the number of species and the fact that, under certain conditions, the thermal pressure ($\tth{p}$) can become negative. The analytic intractability of CDF models and the temperature dependence of Dirac effective mass along with the usage of models with different types of couplings, i.e., density-dependent, non-linear, mixed, nevertheless prevented Ref.~\cite{Raduta_EPJA_2021} from easily tracking the ``ingredients'' which govern the thermal behavior of NM in CDF models, as has been done for the non-relativistic ones~\cite{Constantinou_PRC_2014,Constantinou_PRC_2015}.

The aim of this letter is to bridge this gap and verify numerically some of the assumptions done in the analytical calculations in Ref.~\cite{Constantinou_Annals_2015}. The thermal behavior is studied for a large sample ($10^5$) of EOS models belonging to run~1 in Ref.~\cite{Beznogov_PRC_2023}.  These models assume only nucleonic degrees of freedom, rely on a modified version of the CDF model with simplified density dependent couplings~\cite{Malik_ApJ_2022} and have been generated within a Bayesian approach.  The latter aspect is essential for a thorough sampling of the huge parameter space of this kind of model. Compliance with most current constraints available for cold matter at various densities and isospin asymmetries is fulfilled by accounting for information from nuclear physics, ab initio calculations of pure neutron matter and observations of NSs, for details see Ref.~\cite{Beznogov_PRC_2023}. Throughout this paper natural units will be used ($\hslash=k_\mathrm{B}=c=1$).

\section{The model}
\label{sec:setup}
\subsection{CDF}
\label{ssec:cdf}

The theoretical framework implemented in this work is the CDF theory with density dependent couplings~\cite{Typel_NPA_1999,Hofmann_PRC_2001}.  Nucleons are treated as fundamental particles and the interaction is described by the exchange of $\sigma$-, $\omega$- and $\rho$-``mesons''. The name of the meson thereby determines the quantum numbers for the interaction channel. For the sake of brevity in the following we provide only the most relevant information.  Note that, in spite of being omitted in the equations, the contribution of anti-particles is accounted for in the numerical calculations.

At finite temperature ($T$) the scalar and number densities of nucleons ($i=\mathrm{n,p}$) are given by
\begin{align}
  n_i^{\mathrm s}&= \frac{1}{\pi^2} \int_o^{\infty} dk k^2 \frac{ m^*_i}{E_i(k)}
  f_{\mathrm{FD}}\left(E_i(k)-\mu^*_i \right) ,
  \label{eq:ns}\\
  n_i&= \frac{1}{\pi^2} \int_o^{\infty} dk k^2 f_{\mathrm{FD}}\left(E_i(k)-\mu^*_i \right) ,
  \label{eq:n}
\end{align}
where $k$ and $E_i(k)=\sqrt{k^2+m^{* 2}_i}$ stand for the wave number
and kinetic part of the single particle energy, respectively.  The
Dirac effective masses ($m^*_i$) and effective chemical potentials
($\mu^*_i$) are related to the nucleon mass ($m_{\mathrm N}$)
and the chemical potential ($\mu_i$) via $m^*_i= m_{\mathrm N}- g_{\sigma i}
\bar \sigma$ and $\mu^*_i=\mu_i -g_{\omega i} \bar \omega -g_{\rho i}
t_{3 i} \bar \rho -\Sigma_R$, where $\bar M$ and $g_{M i}$ represent
the mean field expectation value of the corresponding field $M=\sigma,
\omega, \rho$ and its coupling to the nucleon $i$; $t_{3i}$ represents
the third component of isospin of species $i$ with the convention that
$t_{3{\mathrm{p}}}=1/2$; $\Sigma_R$ is the ``rearrangement'' term;
$f_{\mathrm{FD}}\left(x\right)=1/\left[ 1+\exp\left(x/T\right)\right]$
represents the Fermi-Dirac distribution function.

The energy density and pressure can be cast as sums of a kinetic term,
an interaction term and, in the case of pressure, also a ``rearrangement'' term,
\begin{eqnarray}
  e&=&e_{\mathrm{kin}}+e_{\mathrm{int}},\label{eq:ebar} \\
  p&=&p_{\mathrm{kin}}+p_{\mathrm{int}}+p_{\mathrm{rearrang}}. \label{eq:pbar}
\end{eqnarray}

The kinetic terms in eqs.~(\ref{eq:ebar}) and (\ref{eq:pbar}) account for the kinetic contributions of all species,
\begin{eqnarray}
    e_{\mathrm{kin}}&=&\sum_{i={\mathrm n,p}} \frac{1} {\pi^2} \int_0^{\infty}
  dk k^2 E_i(k)
  f_{\mathrm{FD}}\left(E_i(k)-\mu^*_i \right),
  \label{eq:ekin}
 \\
  p_{\mathrm{kin}}&=&\frac 13 \sum_{i={\mathrm n,p}} \frac{1}{\pi^2} \int_0^{\infty}
  \frac{dk k^4}{E_i(k)}
  f_{\mathrm{FD}}\left(E_i(k)-\mu^*_i \right).
  \label{eq:pkin}
\end{eqnarray}

The interaction terms exclusively depend on mean field expectation values of the mesonic fields
\begin{eqnarray}
  e_{\mathrm{int}}&=&\frac{m_{\sigma}^2}{2}\bar\sigma^2+
  \frac{m_{\omega}^2}{2}\bar\omega^2+
  \frac{m_{\rho}^2}{2}\bar\rho^2,
  \label{eq:emeson}
  \\
  p_{\mathrm{int}}&=&-\frac{m_{\sigma}^2}{2}\bar\sigma^2+
  \frac{m_{\omega}^2}{2}\bar\omega^2+
  \frac{m_{\rho}^2}{2}\bar\rho^2,
  \label{eq:pmeson}
\end{eqnarray}
with $m_M$ representing the mass of meson $M$.

The ``rearrangement" term in eq.~(\ref{eq:pbar}) is given by
\begin{equation}
  p_{\mathrm{rearrang}}=n \Sigma_R= n \sum_{\mathrm{i=n,p}} \left(
  \frac{\partial g_{\omega i}}{\partial n_i} \bar \omega n_i +
  t_{3 i} \frac{\partial g_{\rho i}}{\partial n_i} \bar \rho n_i  -
  \frac{\partial g_{\sigma i}}{\partial n_i} \bar \sigma n_i^s
  \right)
  \label{eq:prearrang}
\end{equation}
and arises from the density dependence of the couplings. It is essential for thermodynamic consistency. 

The mean-field expectation values of the meson fields are given by
\begin{align}   
  m_{\sigma}^2 \bar \sigma=\sum_{i={\mathrm n,p}} g_{\sigma i} n_i^s, ~~
  m_{\omega}^2 \bar \omega=\sum_{i={\mathrm n,p}} g_{\omega i} n_i, ~~
  m_{\rho}^2 \bar \rho=\sum_{i={\mathrm n,p}} g_{\rho i} t_{3i} n_i.
  \label{eq:rho}
\end{align}

Computed from the thermodynamic identity $Ts = e + p - \sum_i \mu_i n_i $, the entropy density
takes the form:
\begin{equation}
  s 
  = \frac{e_{\mathrm{kin}} + p_{\mathrm{kin}} - \sum_{i={\mathrm n,p}} \mu_i^* n_i}{T} ~.
\end{equation}

\subsection{EOS models}
\label{ssec:eosmodels}

\renewcommand{\arraystretch}{1.3}
\setlength{\tabcolsep}{2.pt}
\begin{table}
  \caption{Values of selected nuclear matter (NM) parameters corresponding to the EOS models used in this work.  Provided are the saturation density ($\sat{n}$); energy per nucleon ($\sat{E}$), compression modulus ($\sat{K}$), skewness ($\sat{Q}$) and kurtosis ($\sat{Z}$); the symmetry energy ($\sym{J}$), its slope ($\sym{L}$), compressibility ($\sym{K}$), skewness ($\sym{Q}$) and kurtosis ($\sym{Z}$); Dirac effective mass of the nucleons ($\sat{m}^*$), all values given for symmetric nuclear matter (SNM) at $\sat{n}$.  Columns 3 and 4 list the medians and the 68\% confidence interval (CI) of $10^5$ EOS models in the DDB$^*$ set, which corresponds to run~1 in Ref.~\cite{Beznogov_PRC_2023}.  Characteristics of models with extreme values of $\sat{m}^*$ are itemized on columns 5 and 7, respectively. Column 6 corresponds to a model with an intermediate value of $\sat{m}^*$.
  }
\begin{tabular}{lcccccc}
\toprule
\toprule
Par. & Unit & Median &  68\% CI               &  DDB$^*$min  &  DDB$^*$med  & DDB$^*$max \\
\midrule
$\sat{n}$ & $\mathrm {fm}^{-3}$  &  $0.153$  & $^{+0.0049}_{-0.0049}$ &  $0.147$ & $0.152$ & $0.160$  \\
$\sat{E}$ & MeV                 &  $-16.1$  & $^{+0.2}_{-0.2}$      & $-15.7$ & $-15.9$  & $-15.7$  \\
$\sat{K}$ & MeV                 &  $247$    & $^{+33}_{-28}$       & $344.5$ & $228.7$ & $195.9$  \\ 
$\sat{Q}$ & MeV                 &  $-39.9$  & $^{+160}_{-130}$     & $153.9$ & $-124.5$ & $-268.2$ \\
$\sat{Z}$ & MeV                 &  $1360$   & $^{+410}_{-830}$     & $-3756.3$ & $1258.5$ & $1985.2$ \\
$\sym{J}$ & MeV                 &  $32.1$   & $^{+1.8}_{-1.8}$     & $32.6$  & $33.9$ & $33.7$  \\
$\sym{L}$ & MeV                 &  $42.3$   & $^{+15}_{-13}$       & $47.8$  & $59.8$ & $87.8$   \\
$\sym{K}$ & MeV                 &  $-105$   & $^{+27}_{-24}$       & $-86.1$  & $-129.5$ & $-34.3$  \\
$\sym{Q}$ & MeV                 &  $932$    & $^{+360}_{-420}$      & $605.6$ & $556.1$ & $74.8$   \\
$\sym{Z}$ & MeV                 &  $-6440$  & $^{+3100}_{-3800}$    & $-8225.5$ & $-3087.0$ & $-765.5$ \\
$\sat{m}^*$ & $m_{\mathrm N}$        &  $0.657$  & $^{+0.041}_{-0.045}$ & $0.516$  & $0.648$ & $0.751$ \\
\bottomrule
\bottomrule
\end{tabular}
\label{tab:models}
\end{table}
\renewcommand{\arraystretch}{1.0}

For our analysis, we investigate dense matter at finite temperature within each of the $10^5$ EOS
models in run~1 of Ref.~\cite{Beznogov_PRC_2023}, hereafter dubbed
DDB$^*$, and which has been obtained within a Bayesian approach.

For information, the values of some selected NM parameters corresponding to the set DDB$^*$ of EOS models are provided in Table \ref{tab:models}. The large dispersions in the values of the higher order parameters $Q_i$, $Z_i$ are representative of current uncertainties in the behavior of the EOS at high densities, see, e.g., Refs.~\cite{Margueron_PRC_2018,Thi:2021jhz,Char:2023fue}. Values of NM parameters for three particular EOS models within the set are listed, too. Two of them correspond to EOS models with extreme values of the nucleon Dirac effective mass at $\sat{n}$; the third model corresponds to a EOS model with an intermediate value of $\sat{m}^*$. The criterion by which the first of these models have been selected explains why their values of $\sat{K}$ and $\sym{L}$ fall outside the domains recommended in Refs.~\cite{Oertel_RMP_2017,Margueron_PRC_2018} even if this is of no relevance for the present work. 

\subsection{Sommerfeld pseudo-expansion}
\label{ssec:LowT}

\begin{figure}
 \includegraphics[scale=0.4]{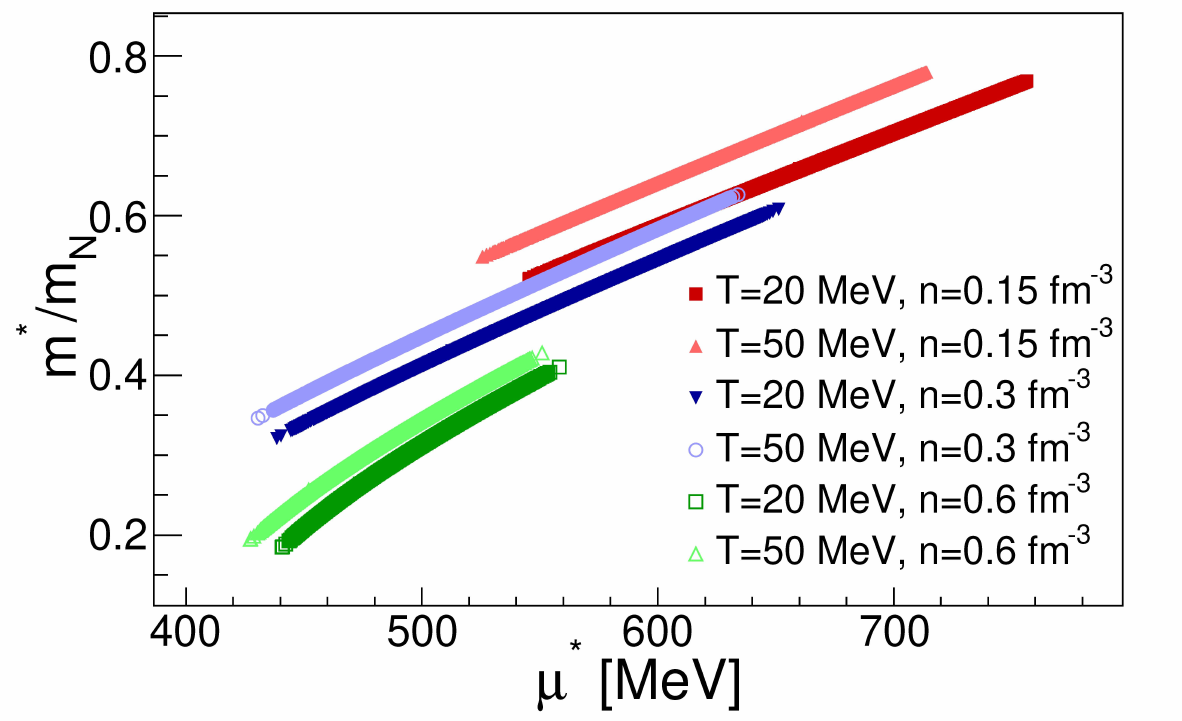}
 \caption{Correlations between the nucleon effective mass ($m^*$) and effective chemical potential ($\mu^*$) in SNM for various thermodynamic conditions. Models correspond to the DDB$^*$ set of effective interactions.
    }
 \label{Fig:meff-mueff}
\end{figure}

\begin{figure}[t]
\includegraphics[width=0.99\columnwidth]{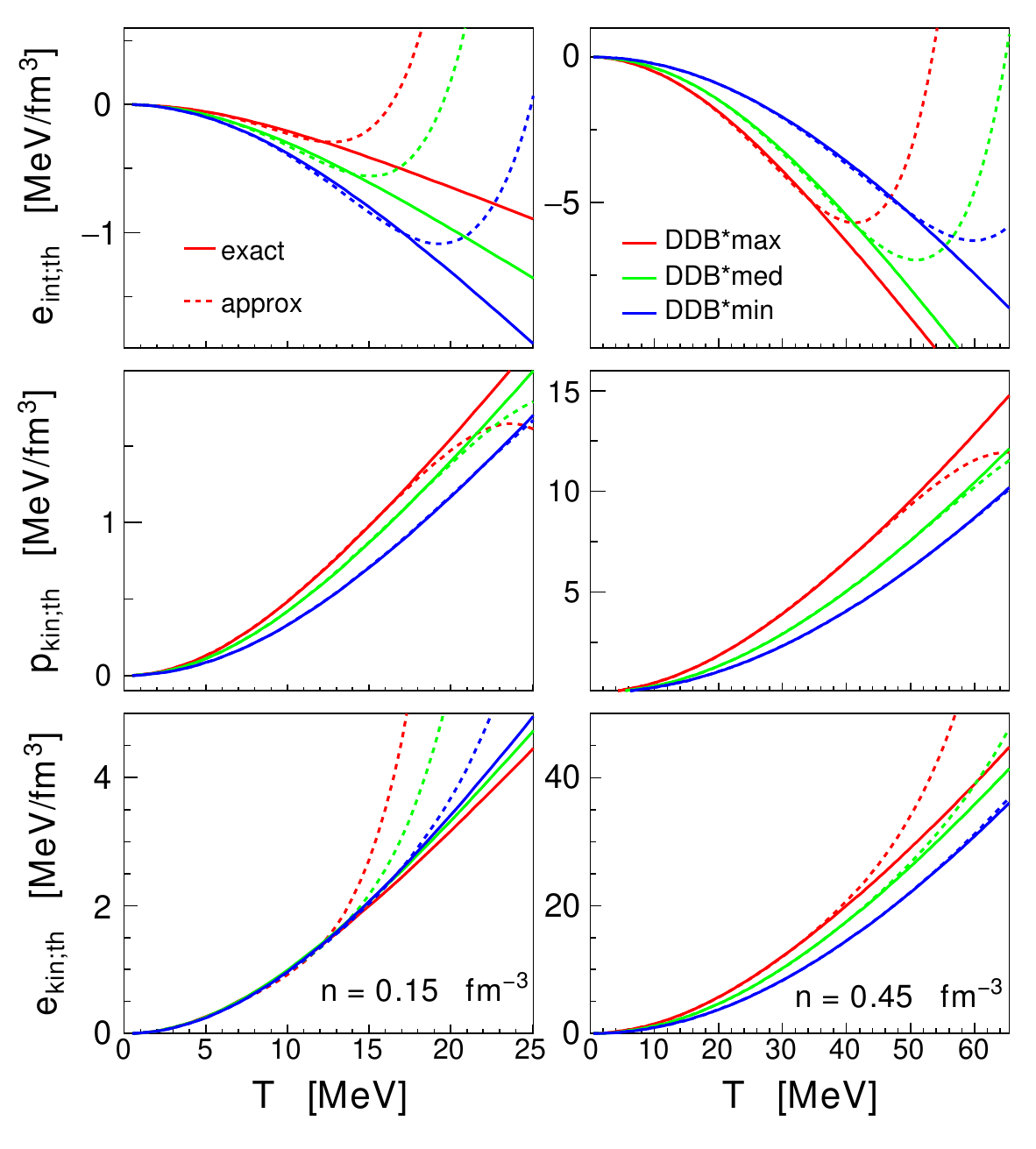}
\caption{
  Test of validity of the low-$T$ approximation. 
  Thermal kinetic energy density ($e_{\mathrm{kin; th}}$), thermal kinetic pressure ($p_{\mathrm{kin; th}}$)  and thermal interaction energy density ($e_{\mathrm{int; th}}$) are represented as functions of temperature in bottom, middle and top panels, respectively. Solid lines are used to illustrate the predictions of eqs.~(\ref{eq:ekin}), (\ref{eq:pkin}) and (\ref{eq:emeson}) with $n^{\mathrm s}_i$ computed according to eq.~(\ref{eq:ns}); predictions of eqs.~(\ref{eq:ekin:lowT}), (\ref{eq:pkin:lowT}) and (\ref{eq:emeson}) with $n^{\mathrm s}_i$ computed according to eq.~(\ref{eq:ns:lowT}) are depicted with dashed lines. Left and right panels correspond to SNM with $n=0.15~\mathrm{fm}^{-3}$ and $n=0.45~\mathrm{fm}^{-3}$, respectively. EOS models under consideration are DDB$^*$min, DDB$^*$med and DDB$^*$max.
  }
\label{fig:test_lowT} 
\end{figure}

In non-relativistic potential models~\cite{Constantinou_PRC_2014}, for a given density and composition, the temperature dependence and, thus, thermal effects arise exclusively from the kinetic contributions. In CDF models, the only explicit $T$-dependence arises from the kinetic contributions, too, but there is an additional implicit temperature dependence via the effective masses, the scalar mesonic field in the interaction terms and, for pressure, the ``rearrangement'' term. All these latter quantities thus contribute to the thermal effects, too. 

In order to understand these effects in the different EOS related quantities, we will now consider a low temperature expansion of the EOS.
At low temperatures ($T \ll \mu^* - m^*$) integrals involving the Fermi-Dirac distribution function can be cast into a series  of even powers of $T$ using the so-called Sommerfeld expansion. In the case of CDF models, this expansion has the generic form:
\begin{align*}  
    I &= \int_0^\infty G(e) f_{\mathrm{FD}}\left(e-\mu^* \right) de  \\
    &\approx \int_0^{\mu^*} G(e) de + \frac{\pi^2}{6} T^2 \left. \frac{dG}{de} \right|_{e=\mu^*} +\frac{7 \pi^4}{360} T^4  \left. \frac{d^3G}{de^3} \right|_{e=\mu^*}+...~,
\end{align*}
where $G$ depends on $m^*$ and both $m^*$ and $\mu^*$ depend on temperature. This means that in this limit all quantities whose structure fits the one in the equation above can be expressed in terms of $m^*$, $\mu^*$ and $T$ only.
We emphasize that due to the $T$-dependence of $m^*$ and $\mu^*$ our following discussion regarding the terms of different orders in temperature is not strictly accurate as the $T$-dependence of $m^*$ and $\mu^*$ is not considered in the expansion (hence, pseudo-expansion).

The expressions for the resulting low $T$ expansions for particle number density, scalar density, density of kinetic energy, kinetic pressure, entropy density and specific heat within CDF models are provided in Appendix A.

Considering now that $m^*$ and $\mu^*$ are strongly correlated, see Fig. \ref{Fig:meff-mueff}, thermal effects are governed by only one quantity in addition to $T$ itself, which is conveniently chosen to be $m^*$. This figure also shows that $m^*$ increases with $T$ whereas $\mu^*$ decreases with $T$; $T$-effects on $\mu^*$ and $m^*$ are equally important.

The validity of the low-$T$ approximation is investigated in Fig.~\ref{fig:test_lowT},
where thermal contributions to kinetic energy density, kinetic pressure and interaction energy density are plotted as a function of $T$.  
The results of eqs.~(\ref{eq:ekin:lowT}) and (\ref{eq:pkin:lowT}) are compared with those of
eqs.~(\ref{eq:ekin}) and (\ref{eq:pkin}). In the interaction energy, for a given density  and composition thermal effects arise due to the dependence on the scalar density. We thus show results for $e_{\mathrm{int; th}}$, with $n_i^{\mathrm s}$ provided by eq.~(\ref{eq:ns:lowT}), and those obtained when eq.~(\ref{eq:ns}) is used instead.
For this comparison, we have considered symmetric nuclear matter (SNM) with $n=0.15~\mathrm{fm}^{-3}$ and $n=0.45~\mathrm{fm}^{-3}$,
and show results for the three models in Table~\ref{tab:models}, with minimum, maximum and intermediate values of $\sat{m}^*$.

The following aspects are worth noticing:
i) for fixed $T$, thermal effects depend on $m^*$,
ii) the validity domain of the low-$T$ approximation shrinks with $m^*$,
iii) for a fixed temperature, $m^*$ effects on $e_{\mathrm{int;\,th}}$ are more pronounced than on $e_{\mathrm{kin;\,th}}$,
iv) $e_{\mathrm{kin;\,th}}$ and $p_{\mathrm{kin;\,th}}$ are positive and increase with $T$,
v) $e_{\mathrm{int;\,th}}=m_{\sigma}^2 [\bar \sigma^2(T)-\bar \sigma^2(T=0) ]/2=-p_{\mathrm{int;\,th}}$ is negative and its absolute value increases with $T$,
vi) $|e_{\mathrm{int;\,th}}| < e_{\mathrm{kin;\,th}}$ and the higher the temperature the less important the contribution of interactions,
vii) $p_{\mathrm{int;\,th}} \approx p_{\mathrm{kin;\,th}}$.

These results can be understood considering that the predictions of eqs.~(\ref{eq:ns:lowT})--(\ref{eq:pkin:lowT})
are the result of the interplay among the different terms entering the low-$T$ expressions.
For the thermodynamic conditions considered here the terms without explicit $T$-dependence in $n^{\mathrm s}$, $e_{\mathrm {kin}}$ and
$p_{\mathrm {kin}}$ decrease with $T$ while those $\propto T^2$ increase with $T$.
The terms $\propto T^4$ do not present a universal behavior. In particular, the contribution to $p_{\mathrm {kin}}$ can be increasing or decreasing.
This means that even if each of these terms increases with $m^*$, since they are alternating, their sum does not necessarily increase.
The opposite $T$-dependence of terms without explicit $T$-dependence and $\propto T^2$
in $n^{\mathrm s}$ explains why the $m^*$-dependence of $e_{\mathrm{int;\,th}}$ is not the same at low and high densities.
The reduced validity domain of the low-$T$ approximation in models with large $m^*$ is merely due to the much steeper increase
of the terms $\propto T^2$, especially for $n^{\mathrm s}$ and $e_{\mathrm {kin}}$.
For DDB$^*$med and DDB$^*$max the high temperatures bending of $p_{\mathrm{kin;\,th}}$ computed according to eq.~\eqref{eq:pkin:lowT}
is caused by the decrease with $T$ of the term $\propto T^4$.

\section{Thermal effects on state variables, thermal coefficients and speed of sound}
\label{sec:Result}

\begin{figure*}
 \includegraphics[scale=0.3]{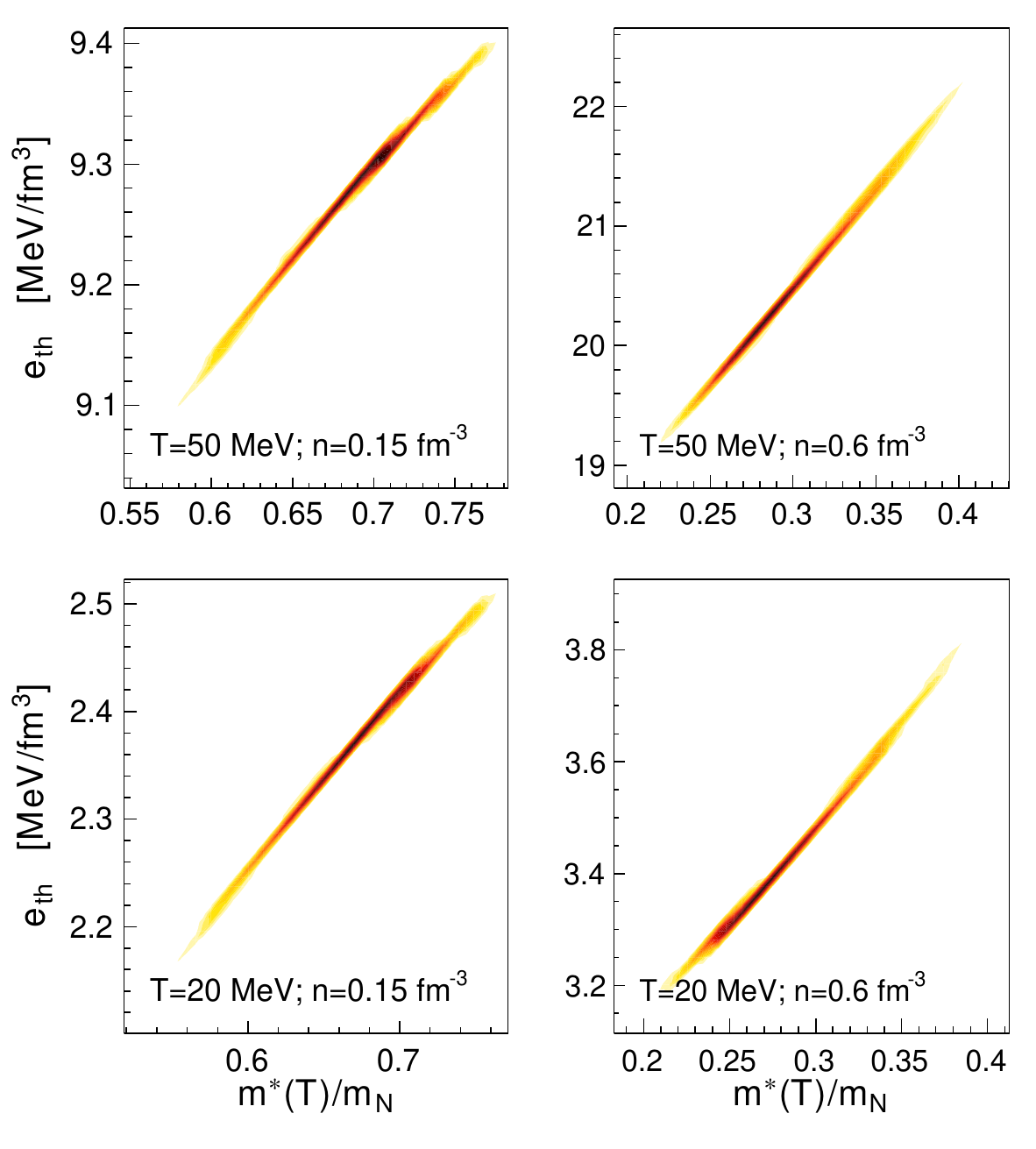}
 \includegraphics[scale=0.3]{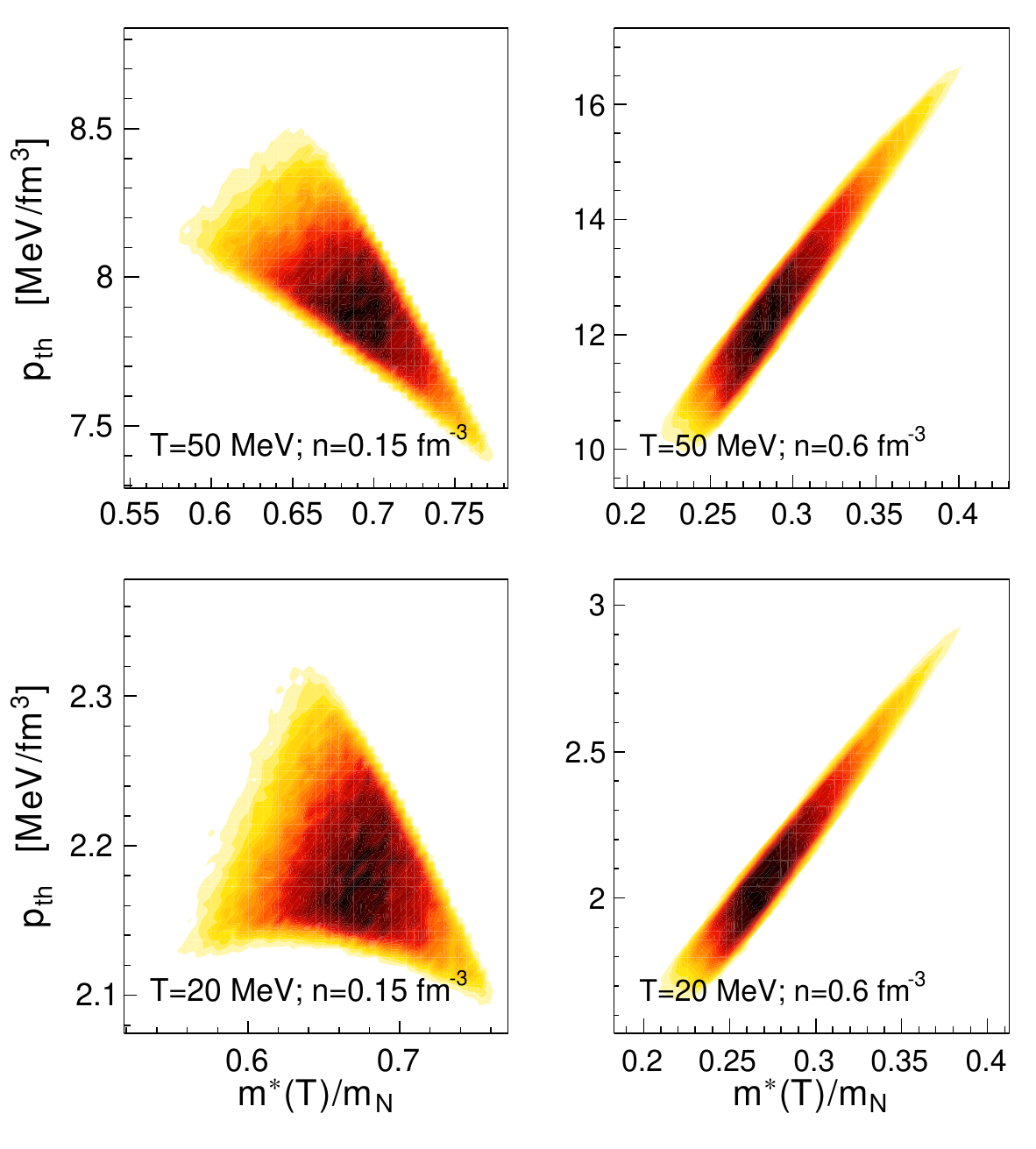}
 \includegraphics[scale=0.3]{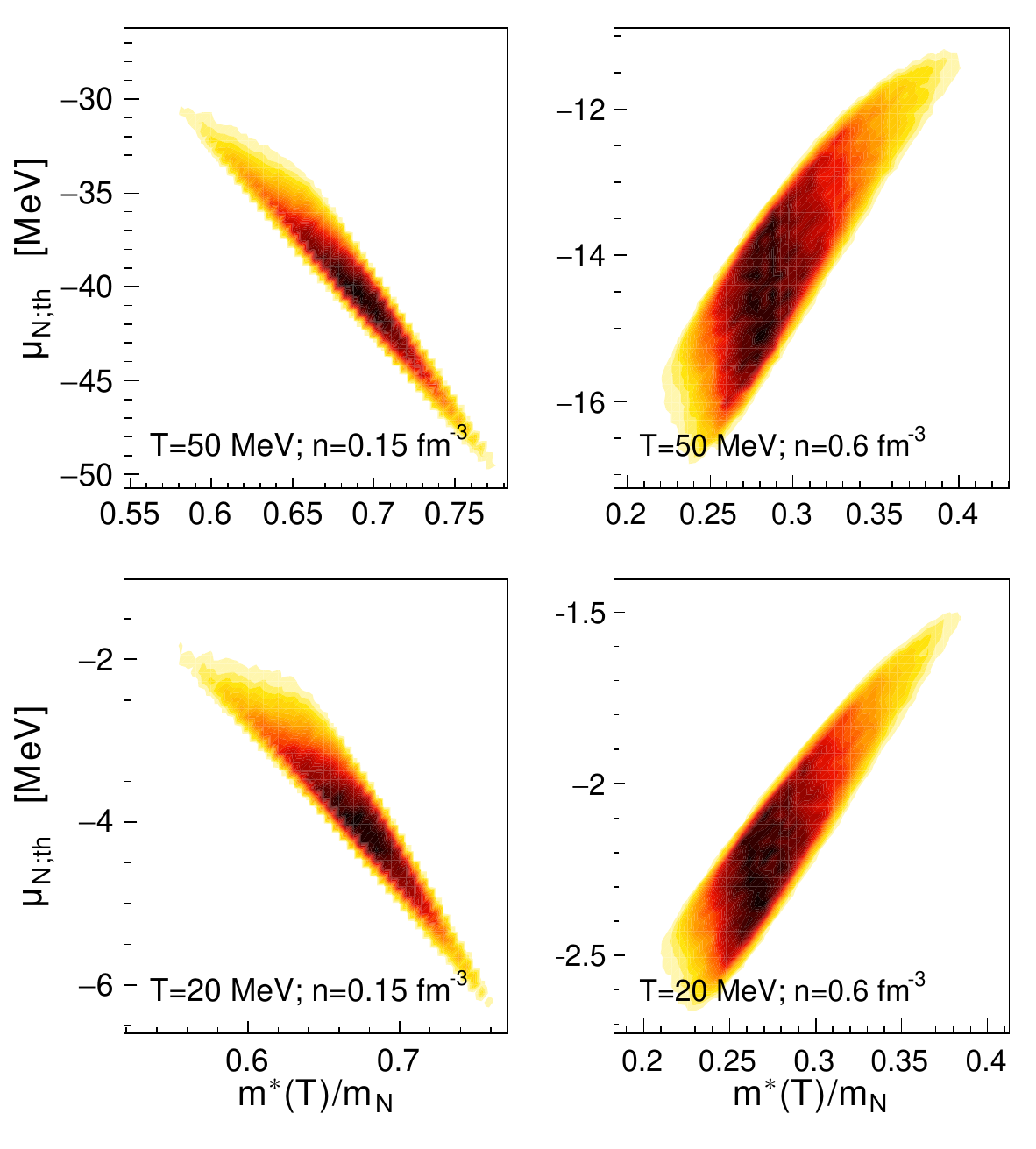}
 \caption{Joint probability density plots for $\tth{e}$, $\tth{p}$ and $\mu_{\mathrm{N; th}}$ vs $m^*$ for the case of SNM under different thermodynamic conditions, as mentioned in each panel. DDB$^*$ set of EOSs.}
 \label{Fig:correl_epmu_m}
\end{figure*}

\begin{figure*}
 \includegraphics[scale=0.3]{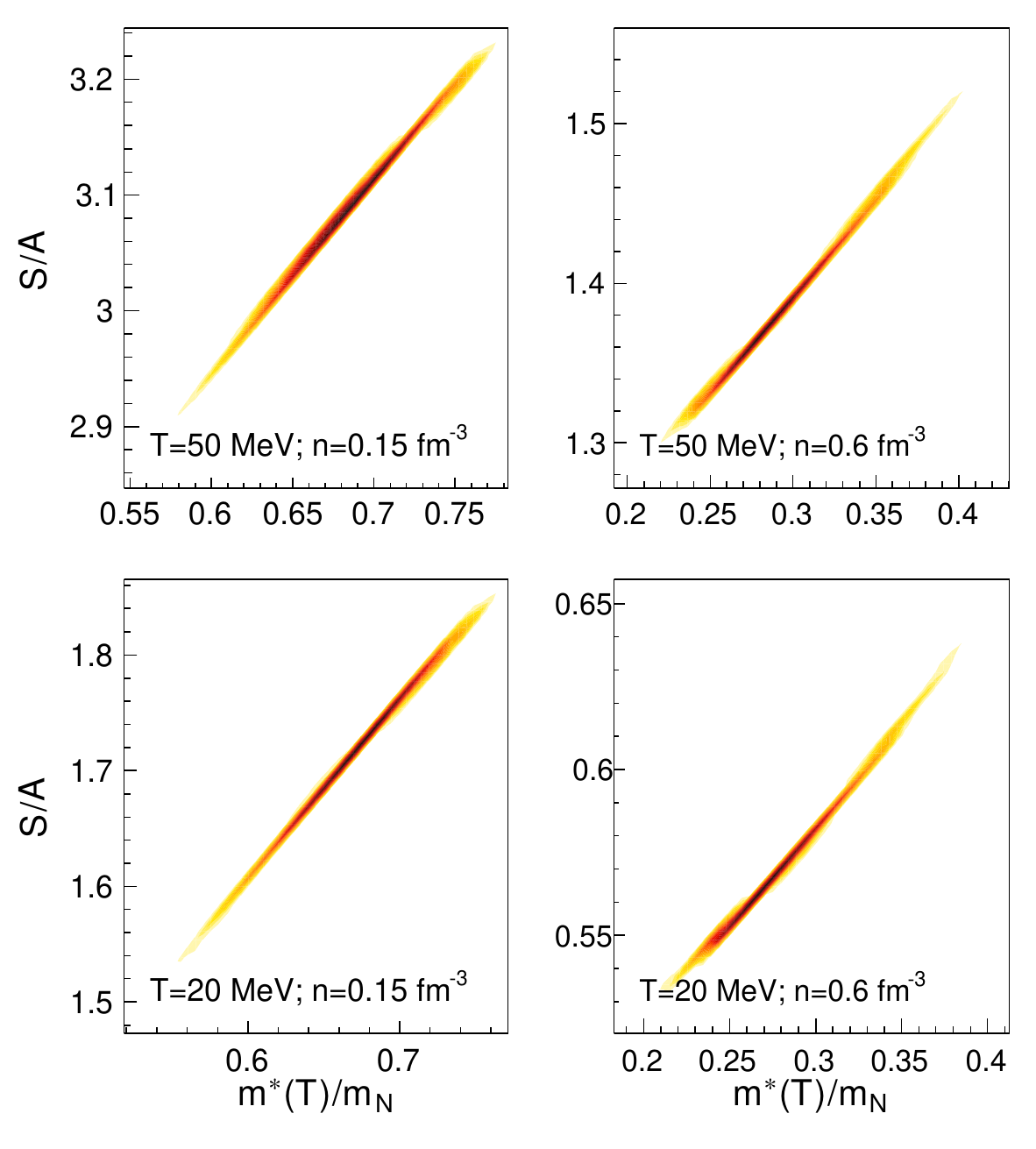}
 \includegraphics[scale=0.3]{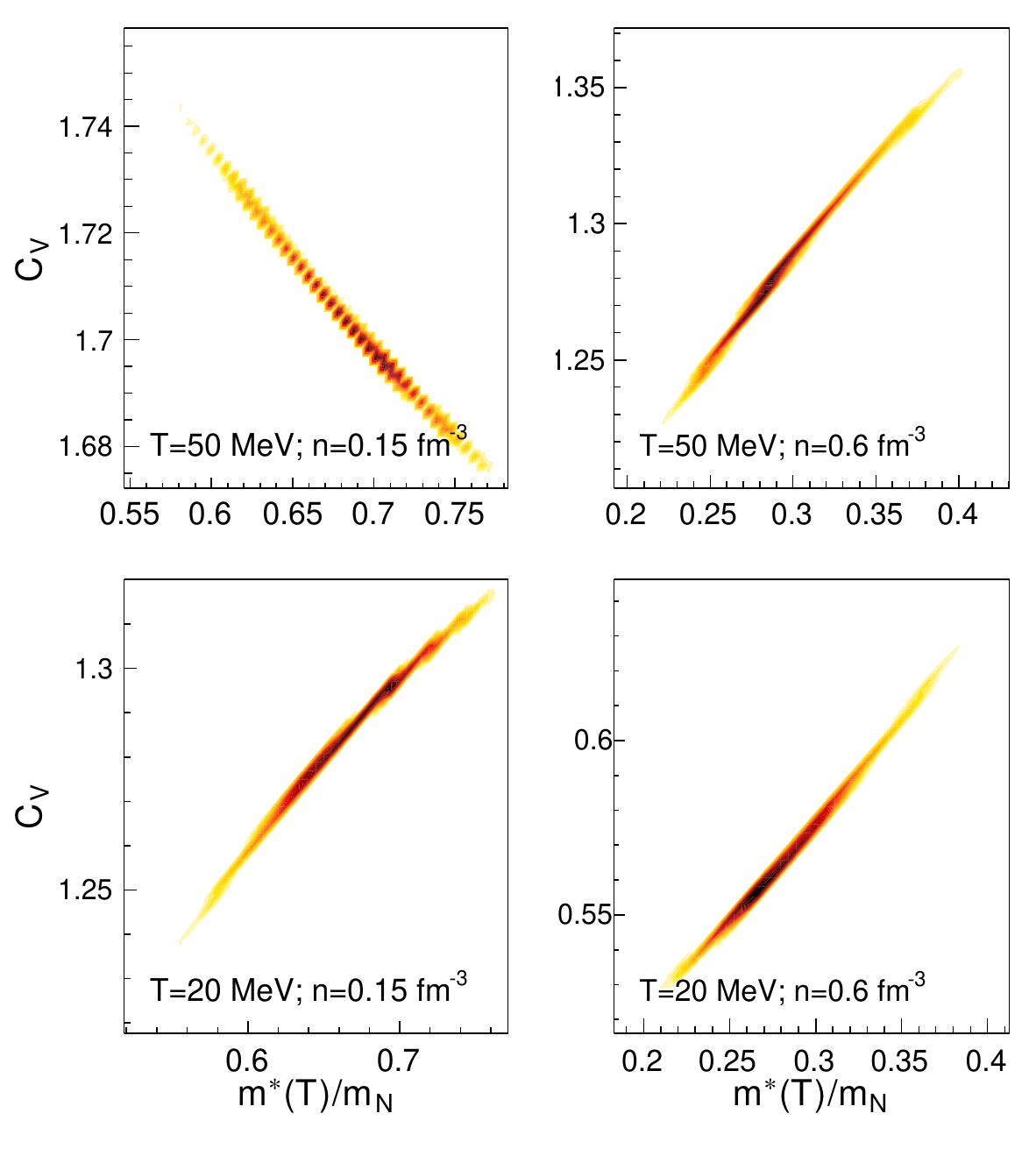}
 \includegraphics[scale=0.3]{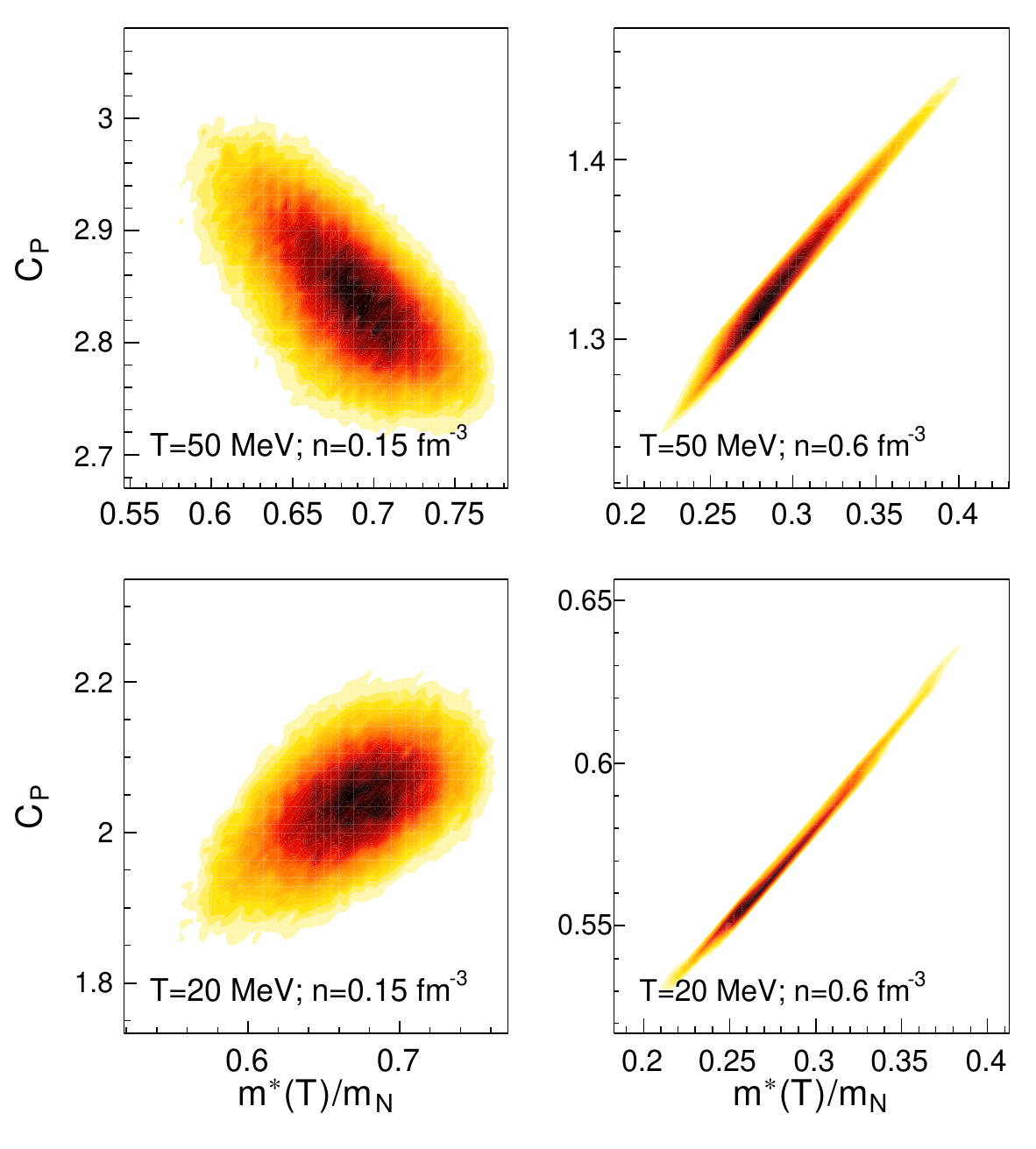}
 \caption{The same as in Fig.~\ref{Fig:correl_epmu_m} but for $S/A$, $C_\mathrm{V}$ and $C_\mathrm{P}$.}
 \label{Fig:correl_SCvCp_m}
\end{figure*}
 
\begin{figure*}
 \includegraphics[scale=0.3]{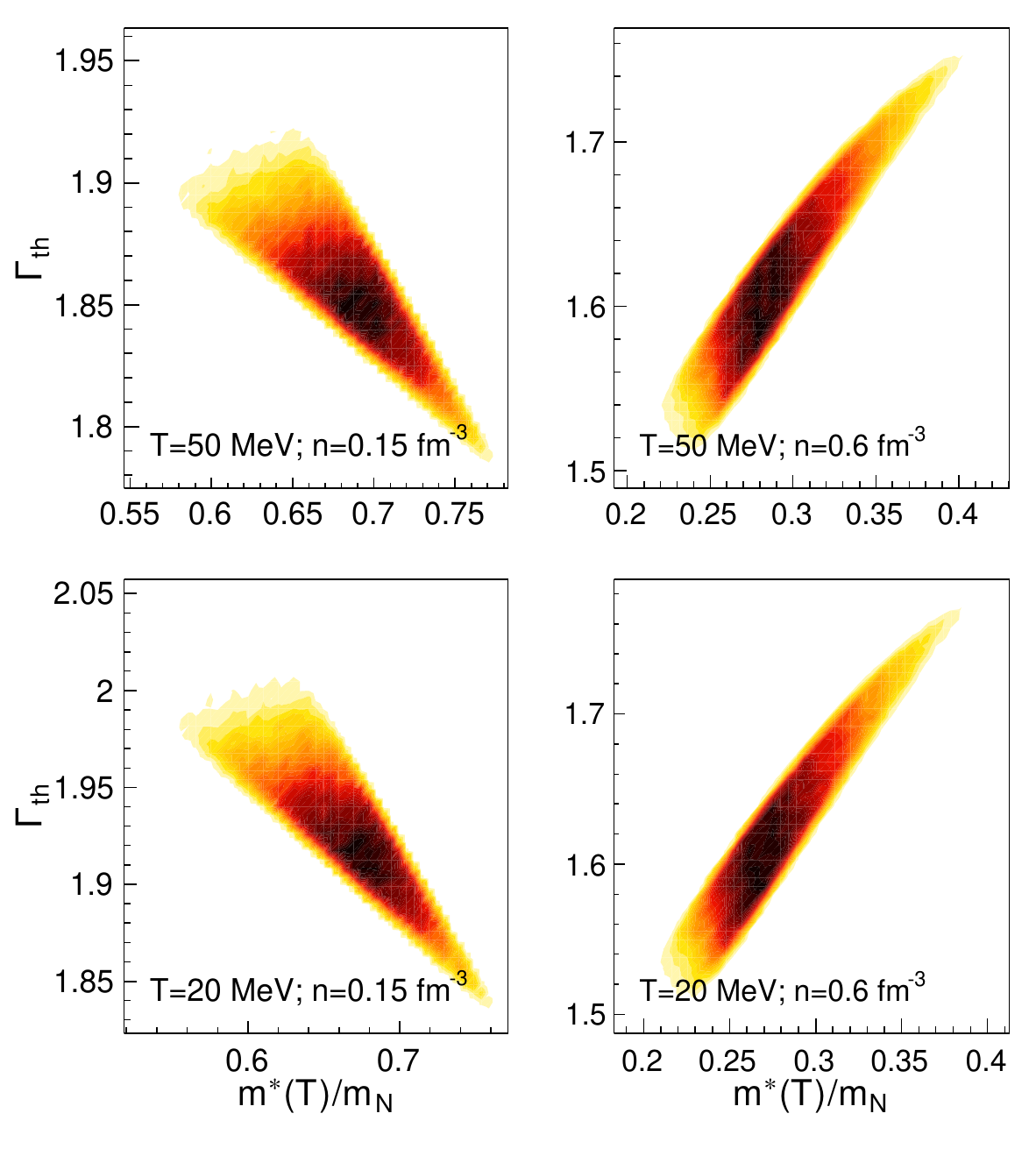}
 \includegraphics[scale=0.3]{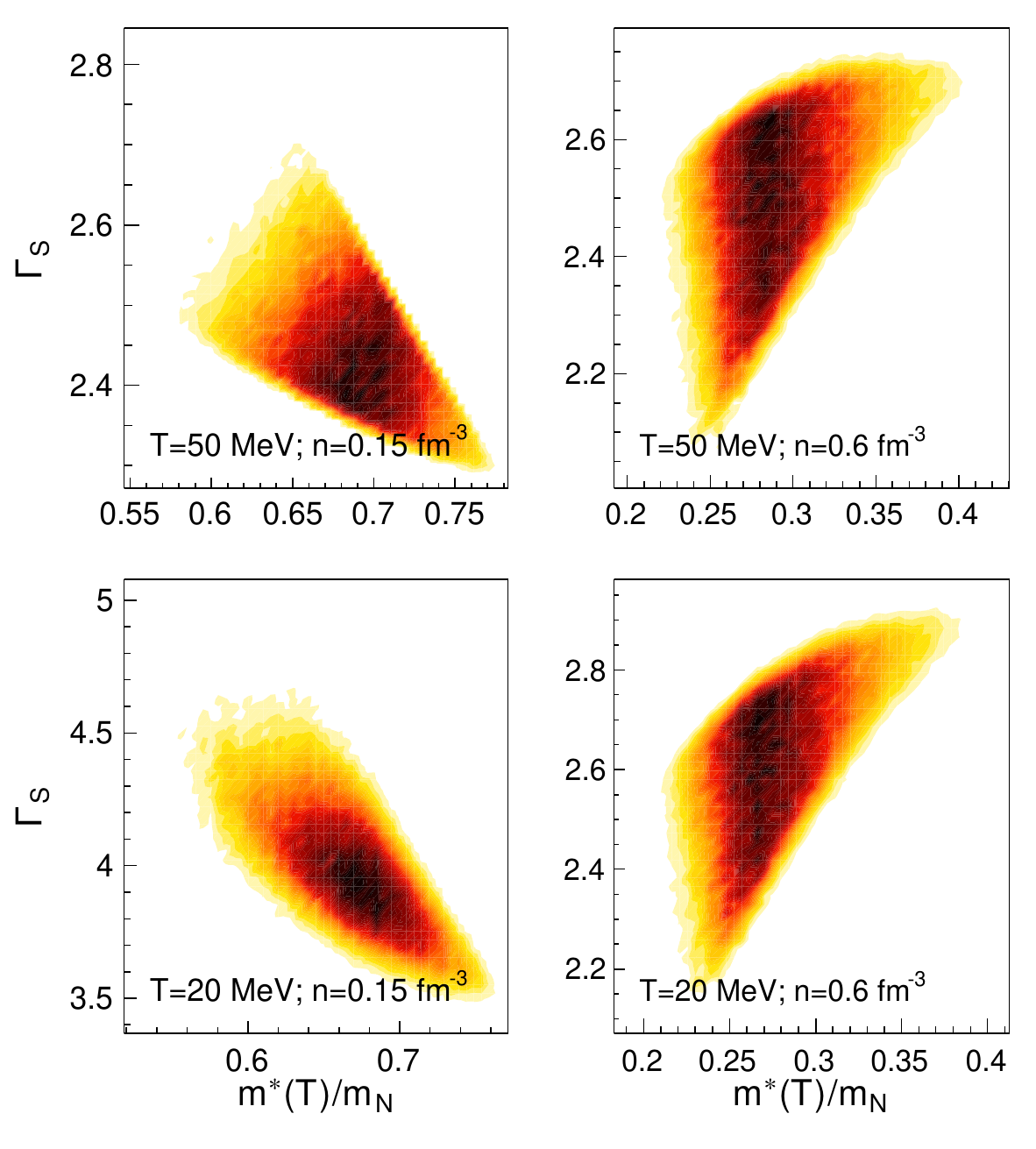}
 \includegraphics[scale=0.3]{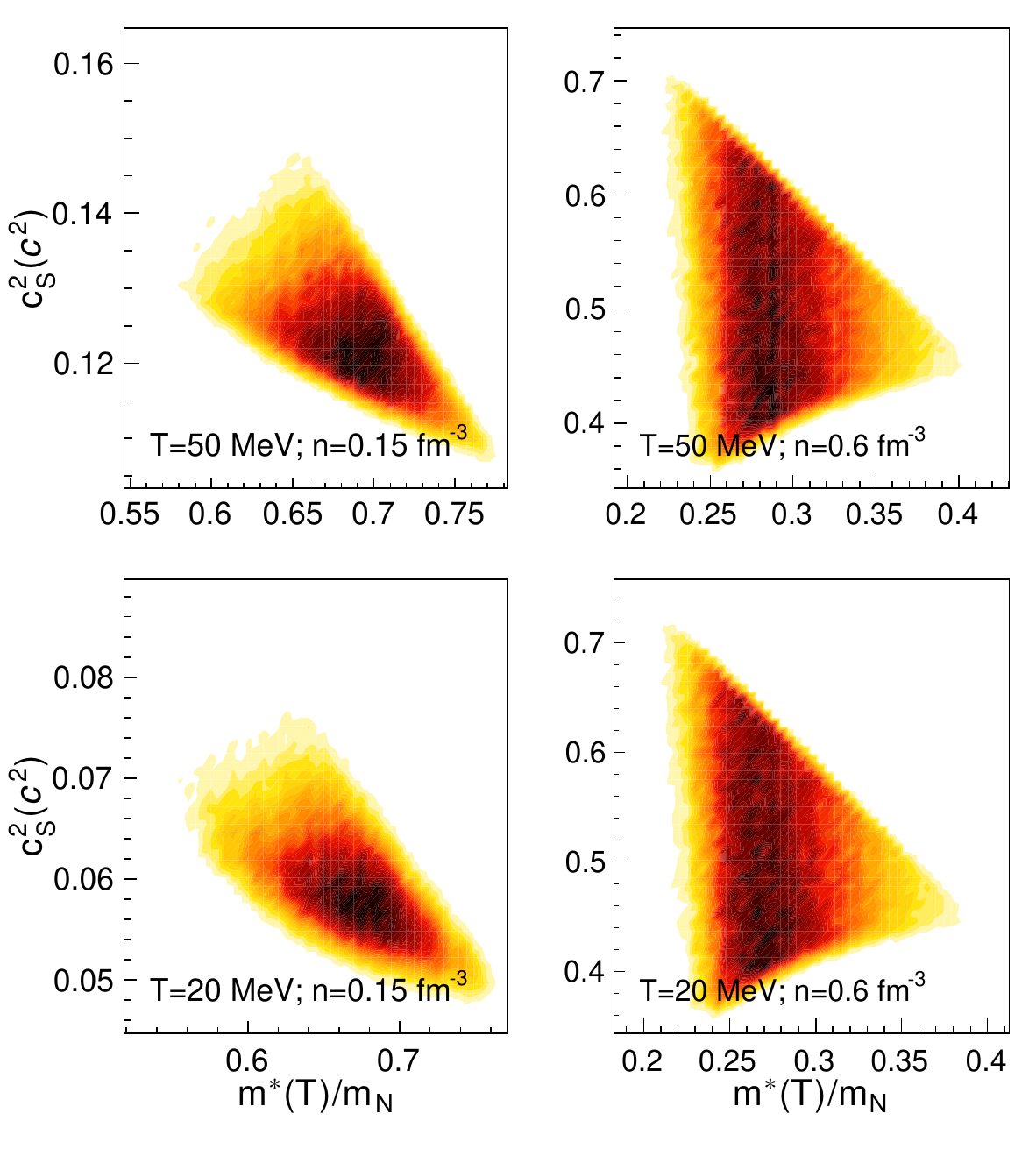}
 \caption{The same as in Fig.~\ref{Fig:correl_epmu_m} but for $\tth{\Gamma}$, $\Gamma_{\mathrm S}$ and $c_\mathrm{S}^2$.}
 \label{Fig:correl_GthGScs2_m}
\end{figure*}
\begin{figure*}
 \includegraphics[scale=0.3]{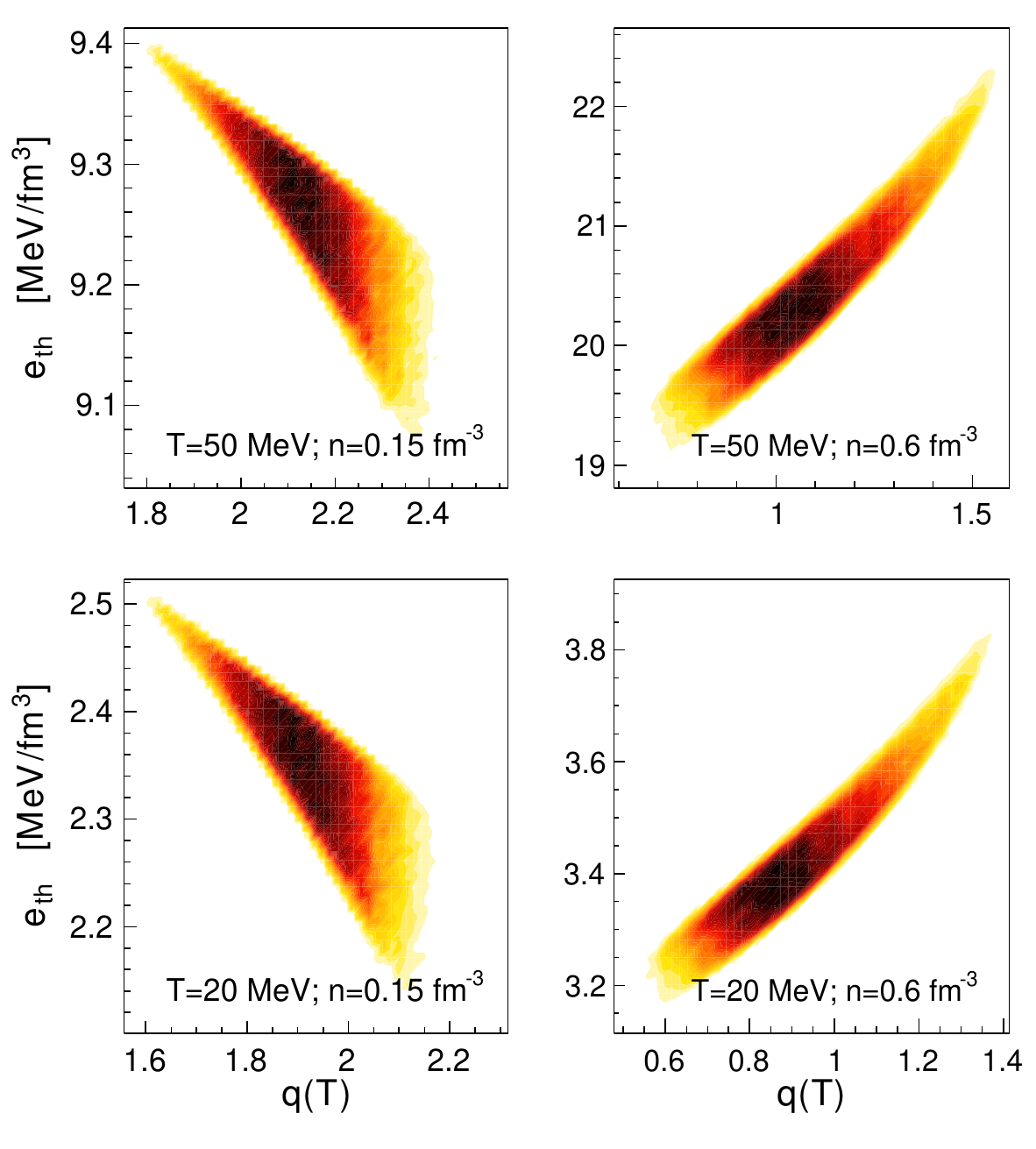}
 \includegraphics[scale=0.3]{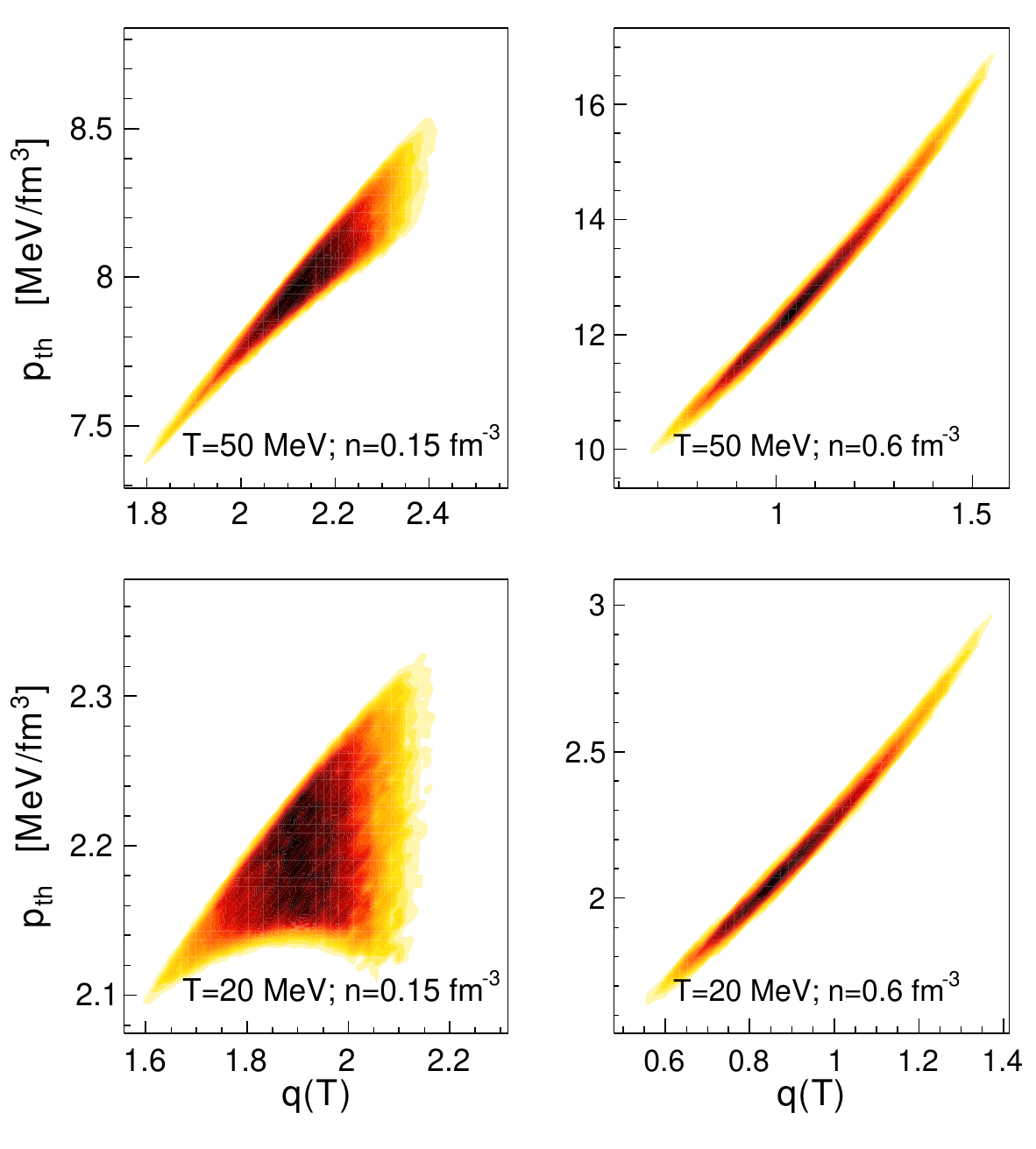}
 \includegraphics[scale=0.3]{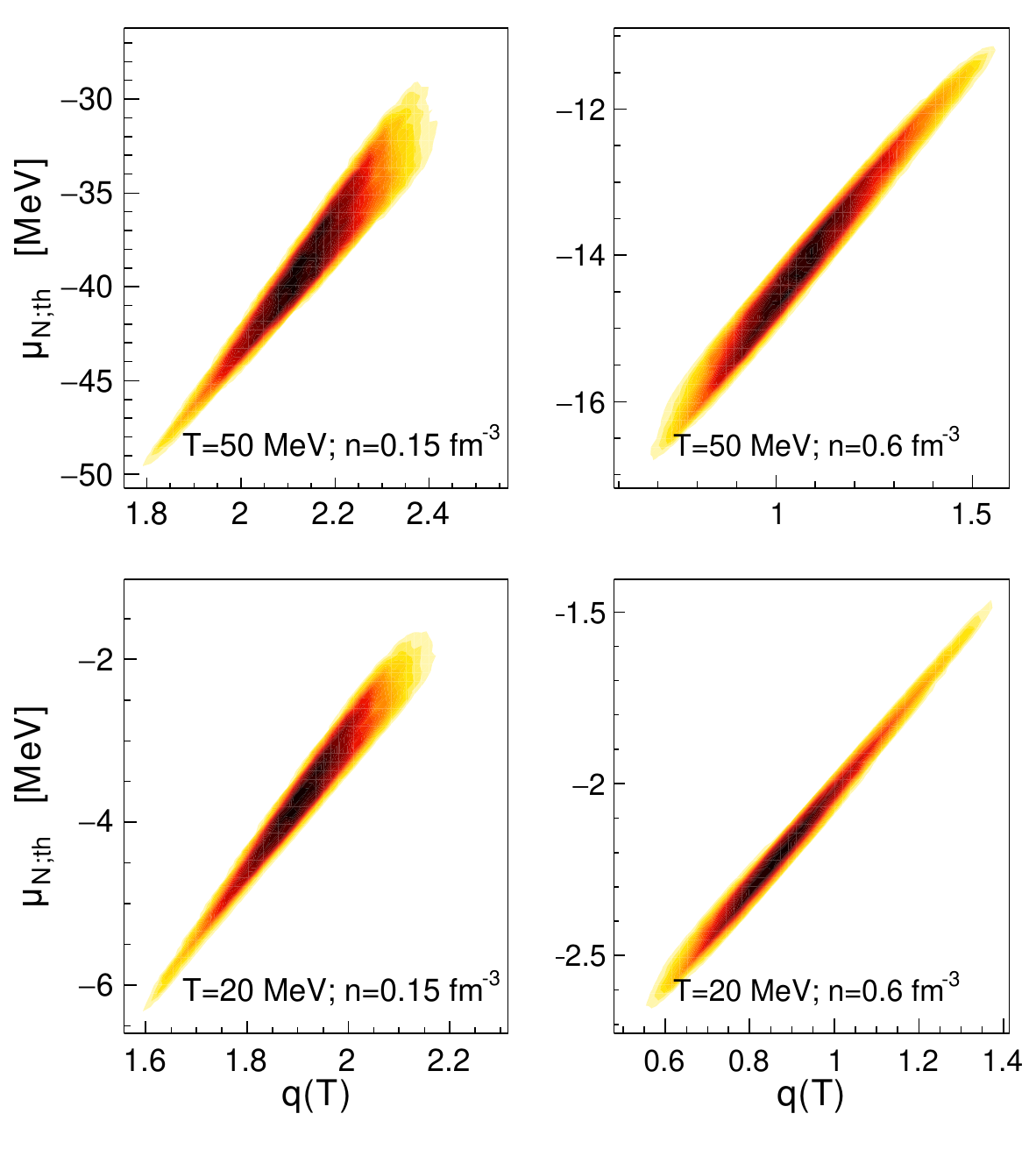}
 \caption{Joint probability density plots for $\tth{e}$, $\tth{p}$ and $\mu_{\mathrm{N; th}}$ vs $q$, eq.~(\ref{eq:q}), for the case of SNM under different thermodynamic conditions, as mentioned in each panel. DDB$^*$ set of EOSs.}
 \label{Fig:correl_epmu_q}
\end{figure*}

\begin{figure*}
 \includegraphics[scale=0.3]{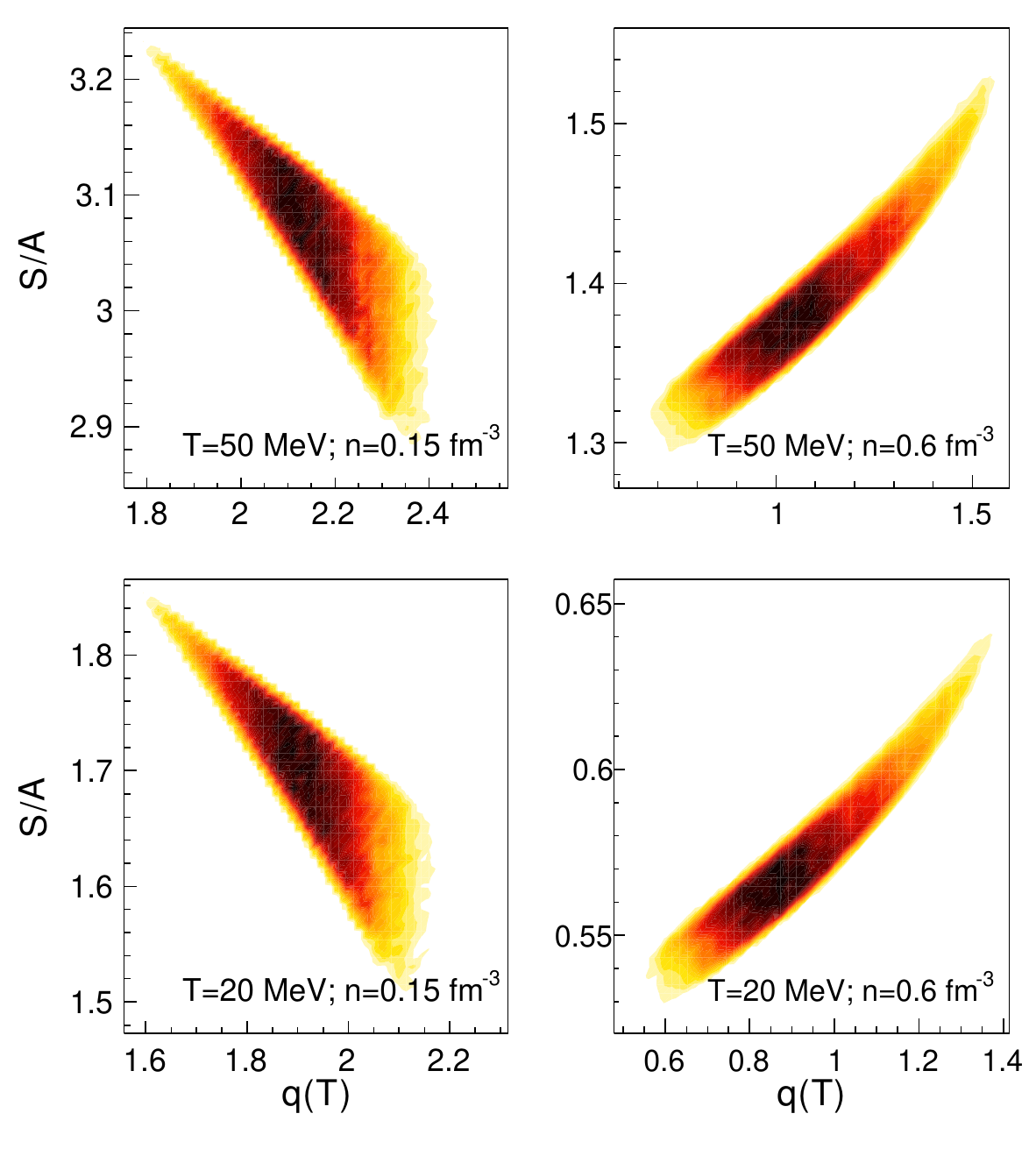}
 \includegraphics[scale=0.3]{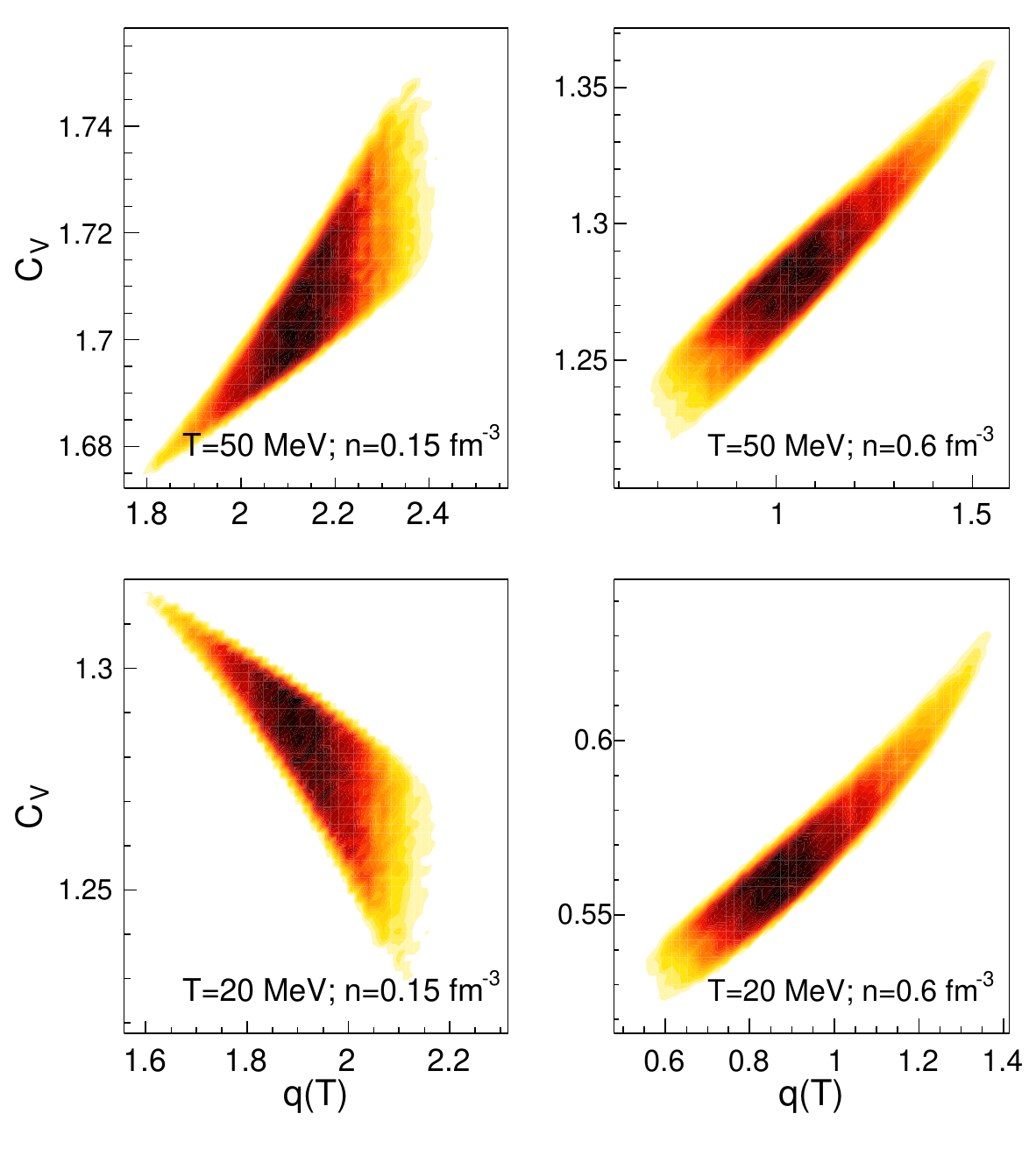}
 \includegraphics[scale=0.3]{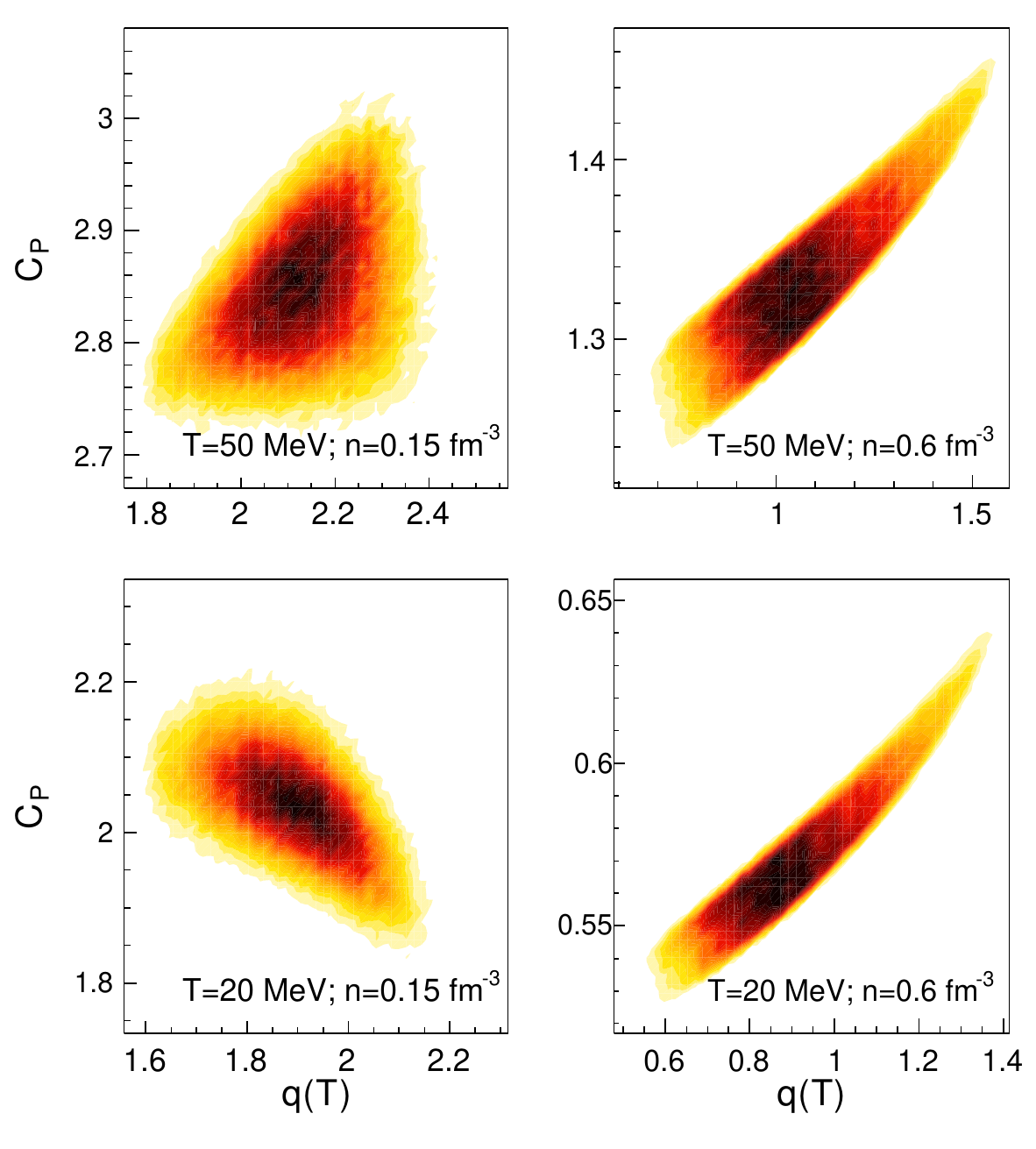}
 \caption{The same as in Fig.~\ref{Fig:correl_epmu_q} but for $S/A$, $C_\mathrm{V}$ and $C_\mathrm{P}$.}
 \label{Fig:correl_SCvCp_q}
\end{figure*}
 
\begin{figure*}
 \includegraphics[scale=0.3]{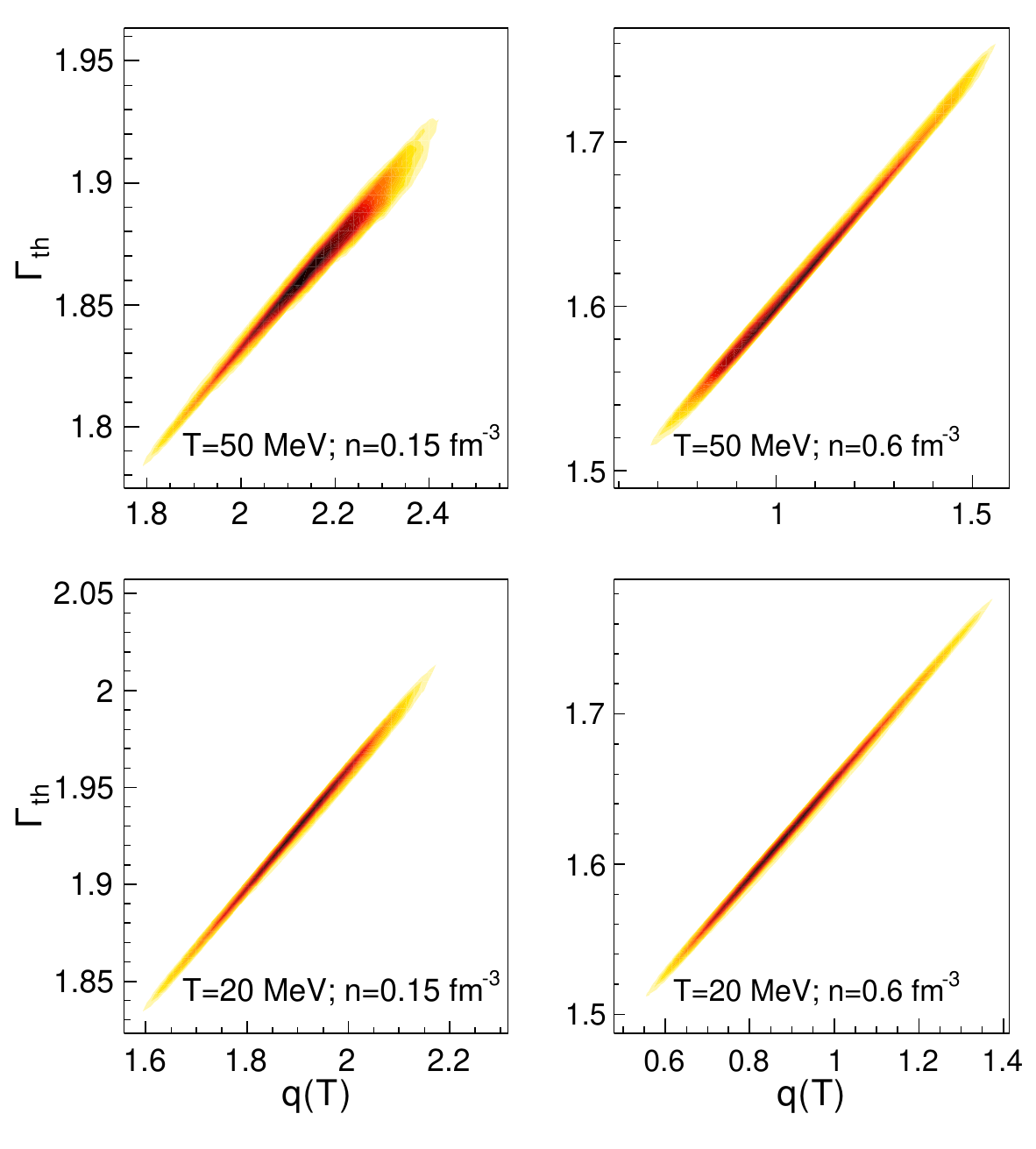}
 \includegraphics[scale=0.3]{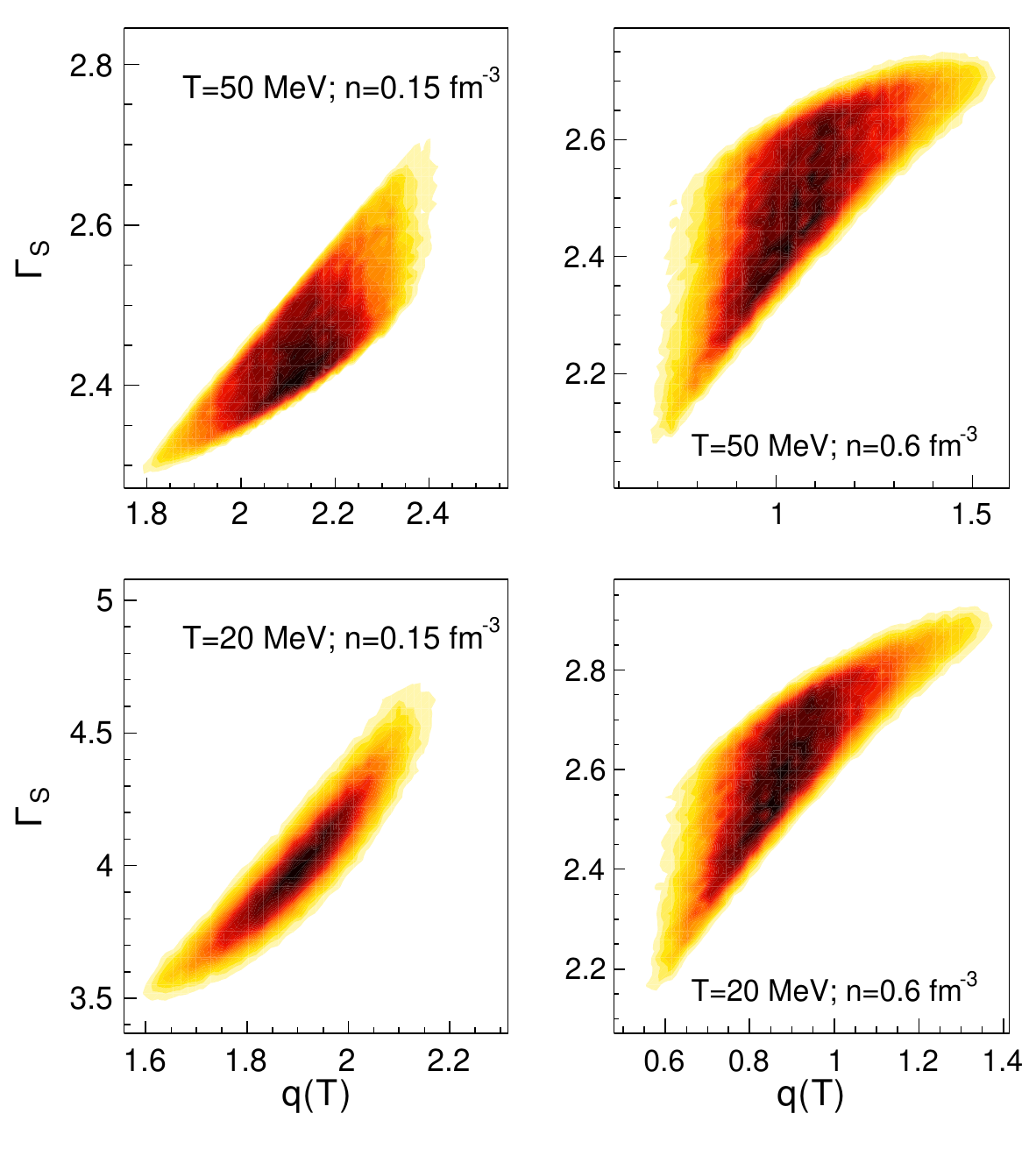}
 \includegraphics[scale=0.3]{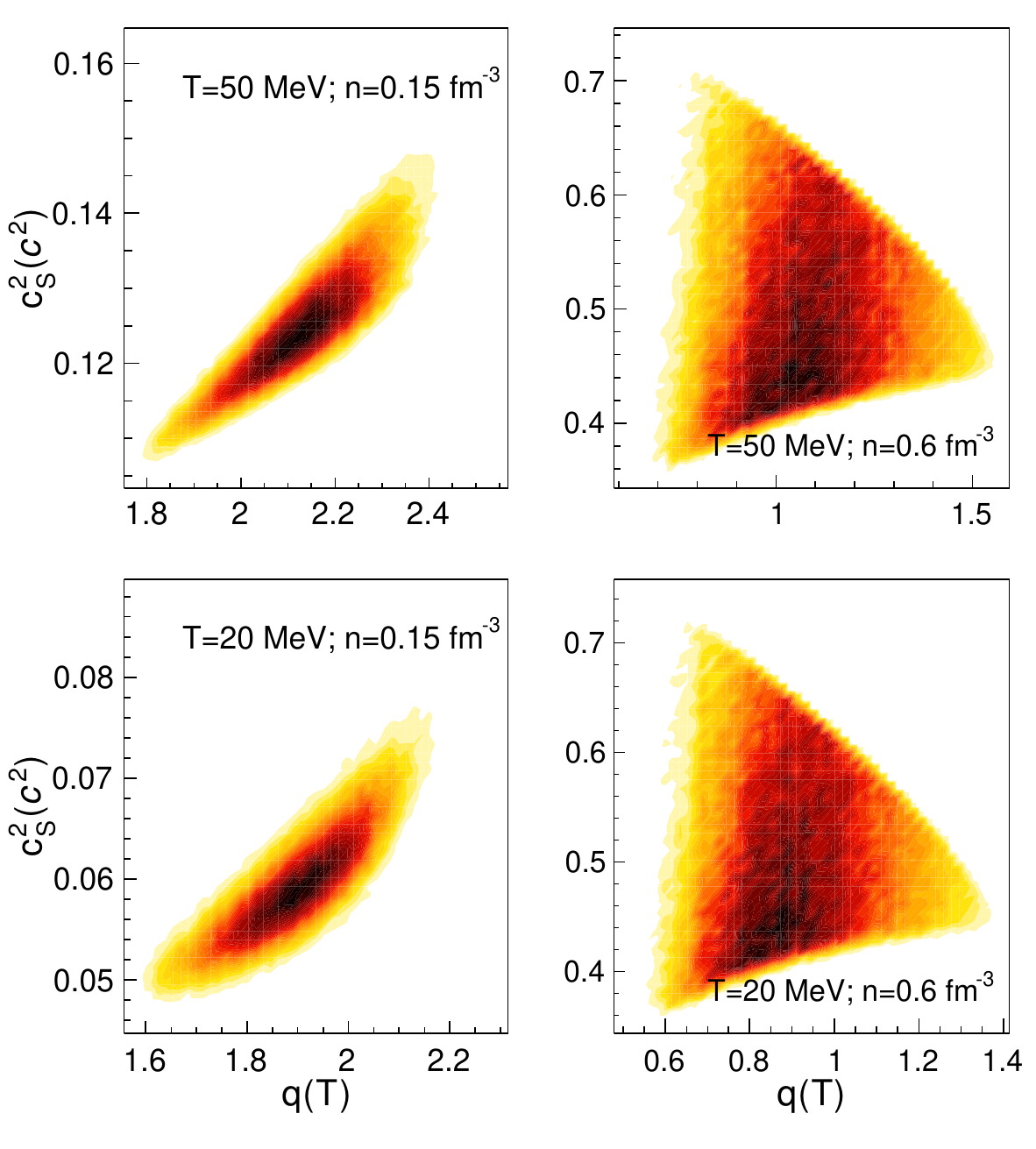}
 \caption{The same as in Fig.~\ref{Fig:correl_epmu_q} but for $\tth{\Gamma}$, $\Gamma_{\mathrm S}$ and $c_\mathrm{S}^2$.}
 \label{Fig:correl_GthGScs2_q}
\end{figure*}

Let us now consider the full set of EOS models and investigate thermal effects for different quantities. As above, only SNM will be considered. Generalization to NM with various degrees of isospin asymmetry, i.e., neutron to proton ratios, is straightforward.

Figs.~\ref{Fig:correl_epmu_m}, \ref{Fig:correl_SCvCp_m} and \ref{Fig:correl_GthGScs2_m} show different thermal quantities as functions of $m^*$ with the aim of identifying potential correlations. Correlations with
\be
q=\frac{m^{*2}}{\mu^{*2}} \left(1-\frac{3n}{m^*} \frac{dm^*}{dn} \right),
\label{eq:q}
\ee
that appears in the Constantinou's et al. next to leading order corrections to thermal state variables~\cite{Constantinou_Annals_2015}, e.g., $\tth{p}$, $\tth{\mu}$, $C_\mathrm{P}$, are examined in Figs.~\ref{Fig:correl_epmu_q}, \ref{Fig:correl_SCvCp_q} and \ref{Fig:correl_GthGScs2_q}. Please note that at variance with Ref.~\cite{Constantinou_Annals_2015}, we consider here the finite-$T$ values of $m^*$ and $\mu^*$. This choice takes into account the small increase of $m^*$ with temperature (at $T=50$~MeV: $\sim$ 2\% to 8\% depending on the model). In particular, we noticed that the stronger increase of $m^*$ with temperature corresponds to the models with low values of $m^*(T=0)$. While not illustrated here, correlations with the zero-temperature values of $m^*$ and $q$ have been analyzed, too, and found to be qualitatively quite similar and of comparable strength as those depicted here. This result is not surprising given the magnitude of the above-mentioned increase of $m^*$ with temperature.
Largely different thermodynamic conditions are considered:
\begin{enumerate}[I]
  \item ($n=0.15~\mathrm{fm}^{-3}$, $T=20~\mathrm{MeV}$)
  \item ($n=0.15~\mathrm{fm}^{-3}$, $T=50~\mathrm{MeV}$)
  \item ($n=0.60~\mathrm{fm}^{-3}$, $T=20~\mathrm{MeV}$)
  \item ($n=0.60~\mathrm{fm}^{-3}$, $T=50~\mathrm{MeV}$)
\end{enumerate}
The third set of conditions (III) thereby lies within the validity domain of the
low-$T$ approximation for all EOS models, whereas the others do not.

The different quantities we are studying here are the thermal energy density ($\tth{e}$), thermal pressure ($\tth{p}$),
thermal chemical potential ($\mu_{\mathrm{th; N}}$), entropy per baryon ($S/A$), thermal index ($\tth{\Gamma}$),
%
heat capacities per nucleon at constant volume
\be
C_\mathrm{V}=T \left.\left(\frac{\partial \left(S/A \right)}{\partial T}\right)\right|_{V,\{ N_i\}},
\label{eq:Cv}
\ee
and at constant pressure
\be
C_p=T \left.\left(\frac{\partial \left(S/A \right)}{\partial T}\right)\right|_{P,\{ N_i\}}
=C_\mathrm{V}+\frac{T}{n^2} \frac{\left.\left( \frac{\partial P}{\partial T}\right|_n\right)^2}{\left.\frac{\partial P}{\partial n}\right|_T}~;
\label{eq:Cp}
\ee
adiabatic index,
\be
\Gamma_\mathrm{S}=\left.\frac{\partial \ln P}{\partial \ln n}\right|_S=
\frac{C_\mathrm{P}}{C_\mathrm{V}} \frac{n}{P} \left.\frac{\partial P}{\partial n}\right|_T~;
\label{eq:GammaS}
\ee
and speed of sound squared (in units of squared speed of light)
\be
c_\mathrm{S}^2=\left.\frac{dP}{de}\right|_{S,A,Y_\mathrm{p}}=\Gamma_\mathrm{S} \frac{P}{e+P}.
\label{eq:cs2}
\ee
In eqs.~(\ref{eq:Cv}) -- (\ref{eq:cs2}) $S$, $V$, $N_i$ and $A$ denote the total entropy, volume,
total number of $i$-particles and total number of particles $A=\sum_i N_i$.

For all thermodynamic conditions, $\tth{e}$, $S/A$ and $C_\mathrm{V}$ perfectly scale with $m^*$. $\tth{e}$ and $S/A$ feature strong, positive and almost linear correlations with $m^*$.
Yet, this almost perfect scaling of $\tth{e}$ with $m^*$ is, in fact, an out-turn of a complicated interplay between $e_{\mathrm{kin;\,th}}$ and $e_{\mathrm{int;\,th}}$. The latter quantities manifest strong correlations with $m^*$, too.
  The sign of these correlations is opposite for $e_{\mathrm{kin;\,th}}$ and $e_{\mathrm{int;\,th}}$ and, moreover, changes with the density.
Within the validity domain of the low-$T$ approximation the behavior of $\tth{e}$, $S/A$ and $C_\mathrm{V}$ is perfectly understandable. The results corresponding to $n=0.15~\mathrm{fm}^{-3}$ show that these correlations survive even beyond the domain of validity of that approximation.
The situation of $C_\mathrm{V}$ is particularly interesting as the shape and sign of the correlation changes.  This is attributable to the behavior of $e_{\mathrm{kin;\,th}}$ and $e_{\mathrm{int;\,th}}$ mentioned above. For condition II, $C_\mathrm{V}$ has values in excess of 1.5. While the same happens in the case of non-relativistic models with finite-range interactions~\cite{Constantinou_PRC_2015}, here the explanation might be different. Condition II approaches the limit of a relativistic ideal gas, for it $C_\mathrm{V}$ can exceed the value of 1.5.
Whenever the degenerate gas limit is approached, i.e., at low temperatures, $S/A \simeq C_\mathrm{V} \simeq C_\mathrm{P}$, which is confirmed for condition III, see Fig.~\ref{Fig:correl_SCvCp_m}.  For large $n$-values, strong and positive correlations between $C_\mathrm{P}$ and $m^*$ occur also at high temperatures. The fact that the correlations between $\tth{p}$ and $m^*$ are weaker than the correlations between $\tth{e}$ and $m^*$ is attributable to the ``rearrangement'' term. Indeed, $p_{\mathrm{kin;\,th}}$ and $p_{\mathrm{int;\,th}}$ show strong correlations with $m^*$ that, as it is the case of $e_{\mathrm{kin;\,th}}$ and $e_{\mathrm{int;\,th}}$, change sign with the density. However, the
extra dependencies introduced by the couplings and their density derivatives result in a ``scattering'' of the $\tth{p}$ dependence on $m^*$.  We note that correlations between $\tth{p}$ and $m^*$ manifest only at high densities.  A sizable dependence of $\mu_{\mathrm{n;th}}$ on $m^*$ manifests as well, though it qualitatively changes with $n$ and $T$.

A number of strong correlations with $q$ occur as well, see Figs.~\ref{Fig:correl_epmu_q}, \ref{Fig:correl_SCvCp_q} and \ref{Fig:correl_GthGScs2_q}. Those involving $\tth{p}$ and $\mu_{\mathrm{th;N}}$ at
($n=0.6~\mathrm{fm}^{-3}$, $T=20~\mathrm{MeV}$) are expected given that in this case the system is close to being degenerate~\cite{Constantinou_Annals_2015}.
In this limit non-negligible positive correlations among $q$ and $\tth{e}$; $S/A$; $C_\mathrm{V}$; $C_\mathrm{P}$ are to be noticed as well.
The remaining panels show that some of those correlations, e.g., $q-\mu_{\mathrm{th;N}}$, persist even outside the validity domain
of the low-$T$ approximation. Particularly interesting is the case of $\tth{\Gamma}$, for which a strong positive linear correlation
with $q$ is obtained under all circumstances. With the exception of ($n=0.15~\mathrm{fm}^{-3}$, $T=20~\mathrm{MeV}$)
this result can be ascribed to the good correlations between $q$ and $\tth{e}$; $\tth{p}$. At $n=0.15~\mathrm{fm}^{-3}$,
a clear correlation exists between $q$ and $c_\mathrm{S}^2$. This means that at finite temperatures the EOS stiffness
around saturation density is regulated by the density dependence of $m^*$.

\section{Summary}
\label{sec:sum}

In this work, we have addressed the EOS dependence of finite-$T$ effects in hot and dense matter built within a CDF framework with
density dependent couplings. $10^5$ EOS models previously generated within a Bayesian approach~\cite{Beznogov_PRC_2023}
to cold NM have been employed. They comply with current constraints from nuclear physics, ab initio calculations of pure neutron matter
and NS observations and ascertain a thorough sampling of the parameter space of this type of EOS models.
Note that other constraints from nuclear structure or heavy ion collisions could be used to constrain cold NM.
While this \emph{might} change the NM and NS properties, we do not expect any strong impact on our finite temperature conclusions.

We found that thermal contributions to state variables and thermal coefficients depend mainly on the Dirac effective mass of the nucleons, with the thermal energy density and entropy per baryon showing very strong correlations. This behavior can be understood within the low-$T$ approximation but extends beyond its validity domain. While the specific heat at constant volume also shows a strong correlation with $m^*$, the sign of this correlation changes with thermodynamic conditions.
Correlations between $\mu_{\mathrm{N;th}}$, $\tth{\Gamma}$, $\Gamma_{\mathrm S}$ and $q$ have been identified as well.
All these correlations consider the values that $m^*$ and $q$ take at finite-$T$. Correlations of similar shapes and strengths have been
found with the $T=0$ values of $m^*$ and $q$. Such correlations have been previously discussed by Constantinou et al.~ \cite{Constantinou_Annals_2015}, who carried out a {\em full} Sommerfeld expansion, i.e., an expansion in actual powers of $T$. Note, however, that to obtain such elegant expressions, Constantinou et al.~\cite{Constantinou_Annals_2015} have assumed that the variation of $m^*$ with temperature can be neglected compared to the variation of $\mu^*$. We have explicitly verified that these variations are, in fact, comparable. Nevertheless, we observed that the correlation patterns are robust with respect to switching between zero and finite temperature values of $m^*$ and $q$. This behavior may be seen as the trivial consequence of the relatively small temperature effects on $m^*$.
From this perspective, our numerical study complements the work done in Ref.~\cite{Constantinou_Annals_2015} and checks its validity
under wider thermodynamic conditions. We expect all the correlations found here to manifest also in other versions of CDF models.
It remains nevertheless to check what exactly determines the magnitude of finite temperature effects and whether
effects stronger than those seen here are realistic.

Numerical simulations of the dynamics of core-collapse supernovae performed with non-relativistic Skyrme-like EOS models showed that the
Landau effective mass governing the thermal effects impacts the contraction of the proto-NS and the neutrino and gravitational wave
signals~\cite{Schneider_PRC_2019b,Yasin_PRL_2020,Andersen_ApJ_2021}, as well as the evolution of proto-NS in failed core-collapse supernovae
and subsequent formation of black holes~\cite{Schneider_ApJ_2020}.
Numerical simulations of the post-merger dynamics of a BNS have been carried out too, establishing links between $m^*$,
the properties of the merger and those of the GW signal~\cite{Fields_ApJL_2023,Raithel_PRD_2023}. These BNS simulations have employed
either Skyrme-like EOS models or a phenomenological model of the density dependence of the effective mass~\cite{Raithel_ApJ_2019}.
The common feature of all these simulations is that they use parameterized effective interactions that allow independent tuning
of various NM parameters and disregard the $T$-dependence of $m^*$.
We conjecture that a similar sensitivity to the Dirac effective mass and its density dependence
will manifest should simulations with EOSs derived within the CDF approach be done.
What remains to be tested is whether the $T$-dependence of $m^*$ enhances the $m^*$ sensitivity of dynamical astrophysics phenomena. 
Confrontation with the results produced by using EOSs derived within dissimilar theoretical frameworks, e.g., relativistic vs
non-relativistic mean field, will be beneficial in many respects. First, this will help to verify the robustness of the finite-$T$
imprints on observational signatures of EOS. Then, it will contribute to assess the uncertainties related to hot matter properties
and their importance with respect to uncertainties related to cold matter, which is an important question to answer in connection
with ongoing planning for future GW detectors.

\nolinenumbers
\section*{Acknowledgements}
A.R.R. and M.V.B. acknowledge financial support from the Ministry of Research, Innovation and Digitization under contracts PN-III-P4-ID-PCE-2020-0293 and PN 23 21 01 02.
M.O. acknowledges financial support from the Agence Nationale de la Recherche (ANR) under contract ANR-22-CE31-0001-01.

\appendix
\section{The low temperature expansion for CDF}
\label{App:LowT}

In this limit, the CDF expressions for particle number density, scalar density, density of kinetic energy, kinetic pressure and entropy density take the forms:
\begin{align}
    \pi^2 n\left(T\right) &=
    \int_0^{\infty} de \sqrt{e^2+2e m^*} \left(e+m^* \right) f_{\mathrm{FD}}\left(e-( \mu^*-m^* )\right) \nonumber \\
    & \approx \frac13 \left( \mu^{* 2}-m^{* 2}\right)^{3/2}+
    \frac{\pi^2 T^2}{6} \frac{2 \mu^{*2} - m^{*2}}{\sqrt{\mu^{*2}-m^{*2}}} 
    \label{eq:n:lowT} \\
    & + \frac{7 \pi^4 T^4}{360} \frac{3m^{*4}}{\left(\mu^{*2}-m^{*2} \right)^{5/2}}~,\nonumber
\end{align}
\begin{align}
    &\frac{\pi^2}{m^*} n^{\mathrm s} \left(T\right)  =
    \int_0^{\infty} de \sqrt{e^2+2e m^*} f_{\mathrm{FD}}\left(e- (\mu^*-m^*) \right) \nonumber \\
    & \approx \frac{\mu^*}{2} \sqrt{ \mu^{*2}-m^{*2}}-\frac{m^{*2}}{2} \ln \left(\frac{\sqrt{ \mu^{*2}-m^{*2}}}{m^*} +\frac{\mu^*}{m^*}\right) \label{eq:ns:lowT} \\
    & +  \frac{\pi^2 T^2}{6}  \frac{\mu^*}{\left( {\mu^{*2}-m^{*2}} \right)^{1/2}} +
    \frac{7 \pi^4 T^4}{360}  \frac{3 m^{*2} \mu^*}{\left(\mu^{*2}-m^{*2} \right)^{5/2}}~, \nonumber
\end{align}
\begin{align}
    e_{\mathrm{kin}}\left(T\right) &= 
    \frac{1}{\pi^2} \int_0^{\infty} de \left(e+m^* \right)^2 \sqrt{e^2+2e m^*} f_{\mathrm{FD}}\left(e-( \mu^*-m^* )\right) \nonumber \\
    & \approx  \frac{\mu^*}{8 \pi^2} \left(2 \mu^{*2} - m^{*2}  \right) \sqrt{\mu^{*2}-m^{*2}} \nonumber \\
    &- \frac{m^{*4}}{8\pi^2} \ln \left( \frac{\sqrt{\mu^{*2}-m^{*2}}}{m^*} +\frac{\mu^*}{m^*}\right) \label{eq:ekin:lowT} \\
    &+ \frac{1}{6} T^2 \frac{\mu^{*} \left( 3 \mu^{*2} -2 m^{*2}\right)}{\left(\mu^{*2}-m^{*2} \right)^{1/2}} \nonumber \\
    &+ \frac{7 \pi^2}{120} T^4 \mu^* \frac{2 \mu^{*4}- 5 m^{*2} \mu^{*2}+4 m^{*4}}{\left(\mu^{*2}-m^{*2}\right)^{5/2}}, \nonumber
\end{align}
\begin{align}
    p_{\mathrm{kin}}\left(T\right) &= \frac{1}{3 \pi^2} \int_0^{\infty} de \left(e^2+2e m^* \right)^{3/2} f_{\mathrm{FD}}\left(e-( \mu^*-m^*) \right) \nonumber \\
    & \approx \frac{\mu^{*}}{24 \pi^2} \left(2 \mu^{*2}-5 m^{*2} \right)\sqrt{\mu^{*2}-m^{*2}} \nonumber \\
    &+ \frac{1}{8 \pi^2} m^{*4} \ln \left( \frac{\sqrt{\mu^{*2}-m^{*2}}}{m^*}+\frac{\mu^*}{m^*}\right) \label{eq:pkin:lowT} \\
    &+ \frac{1}{6} T^2 \mu^* \left(\mu^{*2}-m^{*2}\right)^{1/2}\nonumber \\
    &+ \frac{7 \pi^2}{360} T^4 \mu^* \frac{2 \mu^{*2}-3 m^{*2}}{\left(\mu^{*2}-m^{*2}\right)^{3/2}}~, \nonumber
\end{align}
\be
    s = \frac{T}3 \mu^* \sqrt{\mu^{*2}-m^{*2}} +\frac{7 \pi^2 T^3}{90} \mu^* \frac{2 \mu^{*2}-3 m^{*2}}{\left(\mu^{*2}-m^{*2} \right)^{3/2}}~.
\label{eq:s:lowT}
\ee

The first terms in eqs.~(\ref{eq:n:lowT}) and (\ref{eq:ns:lowT}) along with the sums of the first two terms in eqs.~(\ref{eq:ekin:lowT}) and (\ref{eq:pkin:lowT}) have the same functional forms as at zero temperature. The values of $\mu^*$ and $m^*$ are nevertheless different. Indeed, the $T$- and $n$-dependencies of these two quantities have been omitted for notational simplicity.

From eq.~(\ref{eq:s:lowT}) one can compute the specific heat at constant volume,
\begin{align}
    c_\mathrm{V} &= T\left. \frac{\partial s}{\partial T}\right|_{\{N_i\},V} 
    \label{eq:cv:def} \\
    &= \frac{T}3 \mu^* \sqrt{\mu^{*2}-m^{*2}}+\frac{T^2}{3} \left. \frac{\partial \mu^*}{\partial T}\right|_{\{n_i\}}  \sqrt{\mu^{*2}-m^{*2}}\nonumber\\
   &+\frac{T^3 \pi^2 }{90} \mu^* \frac{22 \mu^{*2}- 53 m^{*2}}{\left( \mu^{*2}-m^{*2}\right)^{3/2}}~.
    \label{eq:cv:lowT}
\end{align}
Here $V$, $N_i$ stand for the total volume and total number of particles belonging to the $i$-species.
We see that, up to the lowest order in $T$, $c_\mathrm{V}=s$; this result is identical to the degenerate limit of a Fermi gas. To cast eq.~\eqref{eq:cv:lowT} we have used the relation
\begin{align}
  &m^* \frac{\partial m^*}{\partial T} \left[ \frac{\mu^{*2}-m^{*2}}{\pi^2} - \frac{m^{*2}}{6\left(\mu^{*2}-m^{*2}\right)} T^2 \right]=\nonumber \\
  &\mu^* \frac{\partial \mu^*}{\partial T} \left[\frac{\mu^{*2}-m^{*2}}{\pi^2} + \frac{2 \mu^{*2}-3m^{*2}}{6 \left(\mu^{*2}-m^{*2} \right)} T^2 \right]
  \label{eq:n=const} \\
  &+\frac{T}{3} \left(2 \mu^{*2}-m^{*2} \right) + \frac{7 \pi^2}{30} T^3 \frac{m^{*4}}{\left(\mu^{*2}-m^{*2} \right)^2}~, \nonumber
\end{align}
which was obtained from eq.~\eqref{eq:n:lowT} by requiring that $\partial n/\partial T = 0$ and keeping terms up to the third order in $T$.

\bibliographystyle{elsarticle-num}
\bibliography{MCMC.bib}

\end{document}